\newcount\mgnf\newcount\tipi\newcount\tipoformule
\newcount\aux\newcount\driver\newcount\cind\global\newcount\bz
\newcount\tipobib\newcount\stile\newcount\modif\newcount\eng
\newcount\noteno\noteno=1

\newdimen\stdindent\newdimen\bibskip
\newdimen\maxit\maxit=0pt

\stile=0         
\tipobib=1       
\bz=0            
\cind=0          
\mgnf=0          
\tipoformule=0   
\aux=1           
\eng=1


\ifnum\mgnf=0
   \magnification=\magstep0
   \hsize=16truecm\vsize=24truecm\hoffset=-0.5cm\voffset=-1.0truecm
   \parindent=4.pt\stdindent=\parindent\fi
\ifnum\mgnf=1
   \magnification=\magstep1\hoffset=-0.5truecm
   \voffset=-1.0truecm\hsize=16truecm\vsize=24.truecm
   \parindent=6pt
   \lineskip=8pt\lineskiplimit=0.1pt      \parskip=0.1pt plus1pt
   \stdindent=\parindent\fi


\def\fine#1{}
\def\draft#1{\bz=1\ifnum\mgnf=1\baselineskip=22pt 
                              \else\baselineskip=16pt\fi
   \ifnum\stile=0\headline={\hfill DRAFT #1}\fi\raggedbottom
    \setbox150\vbox{\parindent=0pt\centerline{\bf Figures' captions}\*}
    \def\gnuins ##1 ##2 ##3{\gnuinsf {##1} {##2} {##3}}
    \def\gnuin ##1 ##2 ##3 ##4 ##5 ##6{\gnuinf {##1} {##2} {##3} 
                {##4} {##5} {##6}} 
    \def\eqfig##1##2##3##4##5##6{\eqfigf {##1} {##2} {##3} {##4} {##5} {##6}}
    \def\eqfigfor##1##2##3##4##5##6##7
             {\eqfigforf {##1} {##2} {##3} {##4} {##5} {##6} {##7}}
      \def\fine ##1{\vfill\eject
                \def\geq(####1){}
               \unvbox150\vfill\eject\raggedbottom
                \centerline{FIGURES}\unvbox149 ##1}}

\def\large{\draft{}\bz=0\headline={\hfill}}


\newcount\prau

\def\titolo#1{\setbox197\vbox{ 
\leftskip=0pt plus16em \rightskip=\leftskip
\spaceskip=.3333em \xspaceskip=.5em \parfillskip=0pt
\pretolerance=9999  \tolerance=9999
\hyphenpenalty=9999 \exhyphenpenalty=9999
\ftitolo #1}}
\def\abstract#1{\setbox198\vbox{
     \centerline{\vbox{\advance\hsize by -2cm \parindent=0pt\it Absrtact: #1}}}}
\def\parole#1{\setbox195\hbox{
     \centerline{\vbox{\advance\hsize by -2cm \parindent=0pt Keywords: #1.}}}}
\def\autore#1#2{\setbox199\hbox{\unhbox199\ifnum\prau=0 #1%
\else, #1\fi\global\advance\prau by 1$^{\simbau}$}
     \setbox196\vbox {\advance\hsize by -\parindent\copy196$^{\simbau}${#2}}}
\def\prima{\unvbox197\vskip1truecm\centerline{\unhbox199}
     \footnote{}{\unvbox196\unvbox194}\vskip1truecm\unvbox198\vskip1truecm\copy195}
\def\simbau{\ifcase\prau
        \or \dagger \or \ddagger \or * \or \star \or \bullet\fi}
\def\grant#1{\setbox196\vbox {\advance\hsize by -\parindent{#1}\copy196}}


\let\a=\alpha \let\b=\beta  \let\g=\gamma     \let\e=\varepsilon
      \let\l=\lambda
                   
\let\s=\sigma     
  \let\o=\omega 
     
            \let\O=\Omega

\let\ge=\geq
\let\le=\leq


{\count255=\time\divide\count255 by 60 \xdef\oramin{\number\count255}
        \multiply\count255 by-60\advance\count255 by\time
   \xdef\oramin{\oramin:\ifnum\count255<10 0\fi\the\count255}}
\def\ora{\oramin }

\def\data{\number\day/\ifcase\month\or gennaio \or febbraio \or marzo \or
aprile \or maggio \or giugno \or luglio \or agosto \or settembre
\or ottobre \or novembre \or dicembre \fi/\number\year;\ \ora}

\def\date{\number\day/\ifcase\month\or January \or February \or March \or
April \or May \or June \or July \or August \or September
\or October \or November \or December \fi/\number\year;\ \ora}

\setbox200\hbox{$\scriptscriptstyle \ifnum\eng=0 \data \else \date \fi$}


\newcount\pgn \pgn=1
\newcount\firstpage

\def\foglio{\number\numsec:\number\pgn\global\advance\pgn by 1}
\def\foglioa{A\number\numsec:\number\pgn\global\advance\pgn by 1}

\def\pagina{\vfill\eject}
\def\ppagina{\ifodd\pageno\pagina\null\pagina\else\pagina\fi}
\def\ppaginan{\ifodd-\pageno\pagina\null\pagina\else\pagina\fi}

\def\setind{\firstpage=\pageno}
\def\setcap#1{\null\def\titlecap{#1}\global\firstpage=\pageno}
\def\titletesi{Indici critici per sistemi fermionici in una dimensione}

\ifnum\stile=1
  \def\pagenumbers{\headline={%
  \ifnum\pageno=\firstpage\hfil\else%
     \ifodd\pageno\hfill{\sc\titlecap}~~{\bf\folio}%
      \else{\bf\folio}~~{\sc\titletesi}\hfill\fi\fi}
  \footline={\ifnum\bz=0
                   \hfill\else\rlap{\hbox{\copy200}\ $\st[\foglio]$}\hfill\fi}}
  \def\pagenumbersind{\headline={%
  \ifnum\pageno=\firstpage\hfil\else%
    \ifodd\pageno\hfill{\rm\romannumeral-\pageno}%
     \else{\rm\romannumeral-\pageno}\hfill\fi\fi}
  \footline={\ifnum\bz=0
                   \hfill\else\rlap{\hbox{\copy200}\ $\st[\foglio]$}\hfill\fi}}
\else
  \def\pagenumbers{\headline={\hfill}
     \footline={\ifnum\bz=0\hfill\folio\hfill
                \else\rlap{\hbox{\copy200}\ $\st[\foglio]$}
                   \hfill{\rm \folio}\hfill\fi}}
\fi

\pagenumbers

\def\numeropag#1{
   \ifnum #1<0 \romannumeral -#1\else \number #1\fi
   }


\global\newcount\numsec\global\newcount\numfor
\global\newcount\numfig\global\newcount\numpar
\global\newcount\numteo\global\newcount\numlem

\numfig=1\numsec=0

\gdef\profonditastruttura{\dp\strutbox}
\def\senondefinito#1{\expandafter\ifx\csname #1\endcsname\relax}
\def\SIA #1,#2,#3 {\senondefinito{#1#2}%
\expandafter\xdef\csname#1#2\endcsname{#3}\else%
\write16{???? ma #1,#2 e' gia' stato definito !!!!}\fi}
\def\etichetta(#1){(\veroparagrafo.\veraformula)
\SIA e,#1,(\veroparagrafo.\veraformula)
 \global\advance\numfor by 1
\write15{\string\FU (#1){\equ(#1)}}
\9{ \write16{ EQ \equ(#1) == #1  }}}
\def \FU(#1)#2{\SIA fu,#1,#2 }
\def\etichettaa(#1){(A\veroparagrafo.\veraformula)
 \SIA e,#1,(A\veroparagrafo.\veraformula)
 \global\advance\numfor by 1
\write15{\string\FU (#1){\equ(#1)}}
\9{ \write16{ EQ \equ(#1) == #1  }}}
\def \FU(#1)#2{\SIA fu,#1,#2 }
\def\tetichetta(#1){\veroparagrafo.\veroteorema
\SIA e,#1,{\veroparagrafo.\veroteorema}
\global\advance\numteo by1
\write15{\string\FU (#1){\equ(#1)}}%
\9{\write16{ EQ \equ(#1) == #1}}}
\def\tetichettaa(#1){A\veroparagrafo.\veroteorema
\SIA e,#1,{A\veroparagrafo.\veroteorema}
\global\advance\numteo by1
\write15{\string\FU (#1){\equ(#1)}}%
\9{\write16{ EQ \equ(#1) == #1}}}
\def\letichetta(#1){\veroparagrafo.\verolemma
\SIA e,#1,{\veroparagrafo.\verolemma}
\global\advance\numlem by1
\write15{\string\FU (#1){\equ(#1)}}%
\9{\write16{ EQ \equ(#1) == #1}}}
\def\getichetta(#1){
 \SIA e,#1,{\verafigura}
 \global\advance\numfig by 1
\write15{\string\FU (#1){\equ(#1)}}
\9{ \write16{ Fig. \equ(#1) ha simbolo  #1  }}}

\def\veroparagrafo{\number\numsec}\def\veraformula{\number\numfor}
\def\verafigura{\number\numfig}\def\veroteorema{\number\numteo}
\def\verolemma{\number\numlem}

\def\geq(#1){\getichetta(#1)\galato(#1)}
\def\Eq(#1){\eqno{\etichetta(#1)\alato(#1)}}
\def\eq(#1){&\etichetta(#1)\alato(#1)}
\def\Eqa(#1){\eqno{\etichettaa(#1)\alato(#1)}}
\def\eqa(#1){&\etichettaa(#1)\alato(#1)}
\def\teq(#1){\tetichetta(#1)\talato(#1)}
\def\teqa(#1){\tetichettaa(#1)\talato(#1)}
\def\leq(#1){\letichetta(#1)\talato(#1)}

\def\Eqr{\eqno(\veroparagrafo.\veraformula)\advance\numfor by 1}
\def\eqr{&(\veroparagrafo.\veraformula)\advance\numfor by 1}
\def\Eqar{\eqno(A\veroparagrafo.\veraformula)\advance\numfor by 1}
\def\eqar{&(A\veroparagrafo.\veraformula)\advance\numfor by 1}

\def\eqv(#1){\senondefinito{fu#1}$\clubsuit$#1\write16{Manca #1 !}%
\else\csname fu#1\endcsname\fi}
\def\equ(#1){\senondefinito{e#1}\eqv(#1)\else\csname e#1\endcsname\fi}


\newdimen\gwidth

\def\commenta#1{\ifnum\bz=1\strut \vadjust{\kern-\profonditastruttura
 \vtop to \profonditastruttura{\baselineskip
 \profonditastruttura\vss
 \rlap{\kern\hsize\kern0.1truecm
  \vbox{\hsize=1.7truecm\raggedright\nota\noindent #1}}}}\fi}
\def\talato(#1){\ifnum\bz=1\strut \vadjust{\kern-\profonditastruttura
 \vtop to \profonditastruttura{\baselineskip
 \profonditastruttura\vss
 \rlap{\kern-1.2truecm{$\scriptstyle#1$}}}}\fi}
\def\alato(#1){\ifnum\bz=1
 {\vtop to \profonditastruttura{\baselineskip
 \profonditastruttura\vss
 \rlap{\kern-\hsize\kern-1.2truecm{$\scriptstyle#1$}}}}\fi}
\def\galato(#1){\ifnum\bz=1 \gwidth=\hsize 
 {\vtop to \profonditastruttura{\baselineskip
 \profonditastruttura\vss
 \rlap{\kern-\gwidth\kern-1.2truecm{$\scriptstyle#1$}}}}\fi}


\newskip\ttglue

\font\ftitolo=cmbx12 
\font\eighttt=cmtt8 \font\sevenit=cmti7  \font\sevensl=cmsl8
\font\sc=cmcsc10

\font\msytw=msbm9 scaled\magstep1
\font\msytww=msbm7 scaled\magstep1


\def\settepunti{\def\rm{\fam0\sevenrm}
\textfont0=\sevenrm \scriptfont0=\fiverm \scriptscriptfont0=\fiverm
\textfont1=\seveni \scriptfont1=\fivei   \scriptscriptfont1=\fivei
\textfont2=\sevensy \scriptfont2=\fivesy   \scriptscriptfont2=\fivesy
\textfont3=\tenex \scriptfont3=\tenex   \scriptscriptfont3=\tenex
\textfont\itfam=\sevenit  \def\it{\fam\itfam\sevenit}%
\textfont\slfam=\sevensl  \def\sl{\fam\slfam\sevensl}%
\textfont\ttfam=\eighttt  \def\tt{\fam\ttfam\eighttt}
\textfont\bffam=\sevenbf  \scriptfont\bffam=\fivebf
\scriptscriptfont\bffam=\fivebf  \def\bf{\fam\bffam\sevenbf}%
\tt \ttglue=.5em plus.25em minus.15em
\setbox\strutbox=\hbox{\vrule height6.5pt depth1.5pt width0pt}%
\normalbaselineskip=20pt\let\sc=\fiverm \normalbaselines\rm}

\let\nota=\settepunti

\def\Rset{\hbox{\msytw R}} 
 \def\Cset{\hbox{\msytw C}}

 \def\Zset{\hbox{\msytw Z}}
\def\sZset{\hbox{\msytww Z}} 
\def\Tset{\hbox{\msytw T}} \def\sTset{\hbox{\msytww T}}


\font\tenmib=cmmib10
\font\sevenmib=cmmib10 scaled 800

\textfont5=\tenmib  \scriptfont5=\sevenmib  \scriptscriptfont5=\fivei

\mathchardef\aaa= "050B
\mathchardef\xxx= "0518
\mathchardef\oo = "0521
\mathchardef\Dp = "0540
\mathchardef\H  = "0548
\mathchardef\FFF= "0546
\mathchardef\ppp= "0570
\mathchardef\nnn= "0517

\newdimen\xshift \newdimen\xwidth \newdimen\yshift \newdimen\ywidth
\newdimen\laln

\def\ins#1#2#3{\nointerlineskip\vbox to0pt {\kern-#2 \hbox{\kern#1 #3}
\vss}}

\def\eqfig#1#2#3#4#5#6{
\xwidth=#1 \xshift=\hsize \advance\xshift 
by-\xwidth \divide\xshift by 2
\yshift=#2 \divide\yshift by 2
\midinsert
\parindent=0pt
\line{\hglue\xshift \vbox to #2{\vfil 
#3 \includegraphics{#4.ps}
}\hfill}
\nobreak
\*
\didascalia{\geq(#6)#5}\endinsert
}

\def\eqfigf#1#2#3#4#5#6{
\xwidth=#1 \xshift=\hsize \advance\xshift 
by-\xwidth \divide\xshift by 2
\yshift=#2 \divide\yshift by 2
\midinsert
\parindent=0pt
\line{\hglue\xshift \vbox to #2{\vfil 
#3 \includegraphics{#4.ps}
}\hfill}
\nobreak
\*
\didascalia{\geq(#6)#5}\endinsert
\setbox149\vbox{\unvbox149 \*\* \centerline{Fig. \equ(#6)} 
\nobreak
\*
\nobreak
\line{\hglue\xshift \vbox to #2{\vfil 
#3 \includegraphics{#4.ps}
}\hfill}\*}
\setbox150\vbox{\unvbox150 \parindent=0pt\* #5\*}
}

\def\eqfigbis#1#2#3#4#5#6#7{
\xwidth=#1 \multiply\xwidth by 2 
\xshift=\hsize \advance\xshift 
by-\xwidth \divide\xshift by 3
\yshift=#2 \divide\yshift by 2
\ywidth=#2
\line{\hfill
\vbox to \ywidth{\vfil #3 \includegraphics{#4.ps}}
\hglue20pt
\vbox to \ywidth{\vfil \includegraphics{#6.ps} #5}
\hfill\raise\yshift\hbox{#7}}}

\def\dimenfor#1#2{\par\xwidth=#1 \multiply\xwidth by 2 
\xshift=\hsize \advance\xshift 
by-\xwidth \divide\xshift by 3
\divide\xwidth by 2 
\yshift=#2 
\ywidth=#2}

\def\eqfigfor#1#2#3#4#5#6#7{
\midinsert
\parindent=0pt
\hbox to \hsize{\hskip\xshift 
\hbox to \xwidth{\vbox to \ywidth{\vfil#2\includegraphics{#10.ps}}\hfill}%
\hskip\xshift%
\hbox to \xwidth{\vbox to \ywidth{\vfil#3\includegraphics{#11.ps}}\hfill}\hfill}
\nobreak
\line{\hglue\xshift 
\hbox to \xwidth{\vbox to \ywidth{\vfil #4 \includegraphics{#12.ps}}\hfill}%
\hglue\xshift
\hbox to \xwidth{\vbox to\ywidth {\vfil #5 \includegraphics{#13.ps}}\hfill}\hfill}
\nobreak
\*\*
\didascalia{\geq(#7)#6}
\endinsert}

\def\eqfigforf#1#2#3#4#5#6#7{
\midinsert
\parindent=0pt
\hbox to \hsize{\hskip\xshift 
\hbox to \xwidth{\vbox to \ywidth{\vfil#2\includegraphics{#10.ps}}\hfill}%
\hskip\xshift%
\hbox to \xwidth{\vbox to \ywidth{\vfil#3\includegraphics{#11.ps}}\hfill}\hfill}
\nobreak
\line{\hglue\xshift 
\hbox to \xwidth{\vbox to \ywidth{\vfil #4 \includegraphics{#12.ps}}\hfill}%
\hglue\xshift
\hbox to \xwidth{\vbox to\ywidth {\vfil #5 \includegraphics{#13.ps}}\hfill}\hfill}
\nobreak
\*\*
\didascalia{\geq(#7)#6}
\endinsert
\setbox149\vbox{\unvbox149\* \centerline{Fig. \equ(#7)} \nobreak\* \nobreak
\*
\vbox{\hbox to \hsize{\hskip\xshift 
\hbox to \xwidth{\vbox to \ywidth{\vfil#2\includegraphics{#10.ps}}\hfill}%
\hskip\xshift%
\hbox to \xwidth{\vbox to \ywidth{\vfil#3\includegraphics{#11.ps}}\hfill}\hfill}
\nobreak
\line{\hglue\xshift 
\hbox to \xwidth{\vbox to \ywidth{\vfil #4 \includegraphics{#12.ps}}\hfill}%
\hglue\xshift
\hbox to \xwidth{\vbox to\ywidth {\vfil #5 \includegraphics{#13.ps}}\hfill}\hfill}
}\hfill}
\setbox150\vbox{\unvbox150\parindent=0pt\* #6\*}
}

\def\eqfigter#1#2#3#4#5#6#7{
\line{\hglue\xshift 
\vbox to \ywidth{\vfil #1 \includegraphics{#2.ps}}
\hglue30pt
\vbox to \ywidth{\vfil #3 \includegraphics{#4.ps}}\hfill}
\multiply\xshift by 3 \advance\xshift by \xwidth \divide\xshift by 2
\line{\hfill\hbox{#7}}
\line{\hglue\xshift 
\vbox to \ywidth{\vfil #5 \includegraphics{#6.ps}}}}


\def\7{\ifnum\modif=1\write13\else\write12\fi}
\def\8{\write13}


\def\gnuin #1 #2 #3 #4 #5 #6{\midinsert\vbox{\vbox to 260pt{
\hbox to 420pt{
\hbox to 200pt{\hfill\nota (a)\hfill}\hfill
\hbox to 200pt{\hfill\nota (b)\hfill}}
\vbox to 110pt{\vfill\hbox to 420pt{
\hbox to 200pt{\includegraphics{#1.ps}\hfill}\hfill
\hbox to 200pt{\includegraphics{#2.ps}\hfill}
}}\vfill
\hbox to 420pt{
\hbox to 200pt{\hfill\nota (c)\hfill}\hfill
\hbox to 200pt{\hfill\nota (d)\hfill}}
\vbox to 110pt{\vfill\hbox to 420pt{
\hbox to 200pt{\includegraphics{#3.ps}\hfill}\hfill
\hbox to 200pt{\includegraphics{#4.ps}\hfill}
}}\vfill}
\vskip0.25cm
\0\didascalia{\geq(#5): #6}}
\endinsert}

\def\gnuinf #1 #2 #3 #4 #5 #6{\midinsert\nointerlineskip\vbox to 260pt{
\hbox to 420pt{
\hbox to 200pt{\hfill\nota (a)\hfill}\hfill
\hbox to 200pt{\hfill\nota (b)\hfill}}
\vbox to 110pt{\vfill\hbox to 420pt{
\hbox to 200pt{\includegraphics{#1.ps}\hfill}\hfill
\hbox to 200pt{\includegraphics{#2.ps}\hfill}
}}\vfill
\hbox to 420pt{
\hbox to 200pt{\hfill\nota (c)\hfill}\hfill
\hbox to 200pt{\hfill\nota (d)\hfill}}
\vbox to 110pt{\vfill\hbox to 420pt{
\hbox to 200pt{\includegraphics{#3.ps}\hfill}\hfill
\hbox to 200pt{\includegraphics{#4.ps}\hfill}
}}\vfill}
\?
\0\didascalia{\geq(#5): #6}
\endinsert
\global\setbox150\vbox{\unvbox150 \*\*\0 Fig. \equ(#5): #6}
\global\setbox149\vbox{\unvbox149 \*\*
    \vbox{\centerline{Fig. \equ(#5)(a)} \nobreak
    \vbox to 200pt{\vfill\includegraphics{#1.ps_f}}}\*\*
    \vbox{\centerline{Fig. \equ(#5)(b)}\nobreak
    \vbox to 200pt{\vfill\includegraphics{#2.ps_f}}}\*\*
    \vbox{\centerline{Fig. \equ(#5)(c)}\nobreak
    \vbox to 200pt{\vfill\includegraphics{#3.ps_f}}}\*\*
    \vbox{\centerline{Fig. \equ(#5)(d)}\nobreak
    \vbox to 200pt{\vfill\includegraphics{#4.ps_f}}}
}}

\def\gnuins #1 #2 #3{\midinsert\nointerlineskip
\vbox{\line{\vbox to 220pt{\vfill
\includegraphics{#1.ps}}\hfill}
\*
\0\didascalia{\geq(#2): #3}}
\endinsert}

\def\gnuinsf #1 #2 #3{\midinsert\nointerlineskip
\vbox{\line{\vbox to 220pt{\vfill
\includegraphics{#1.ps}}\hfill}
\*
\0\didascalia{\geq(#2): #3}}\endinsert
\global\setbox150\vbox{\unvbox150 \*\*\0 Fig. \equ(#2): #3}
\global\setbox149\vbox{\unvbox149 \*\* 
    \vbox{\centerline{Fig. \equ(#2)}\nobreak
    \vbox to 220pt{\vfill\includegraphics{#1.ps_f}}}} 
}


\def\9#1{\ifnum\aux=1#1\else\relax\fi}
\let\numero=\number
\def\boh{\hbox{$\clubsuit$}\write16{Qualcosa di indefinito a pag. \the\pageno}}
\def\didascalia#1{\vbox{\nota\0#1\hfill}\vskip0.3truecm}
\def\frac#1#2{{#1\over #2}}
\def\V#1{\underline{#1}}
       
          \let\i=\infty
            
        \let\0=\noindent
\def\guida{\leaders\hbox to 1em{\hss.\hss}\hfill}
\def\tende#1{\vtop{\ialign{##\crcr\rightarrowfill\crcr
              \noalign{\kern-1pt\nointerlineskip}
              \hglue3.pt${\scriptstyle #1}$\hglue3.pt\crcr}}}
\def\otto{{\kern-1.truept\leftarrow\kern-5.truept\to\kern-1.truept}}

\def\={{ \; \equiv \; }}             
\ifnum\mgnf=0
    \def\openone{\leavevmode\hbox{\ninerm 1\kern-3.3pt\tenrm1}}%
\fi
\ifnum\mgnf=1
     \def\openone{\leavevmode\hbox{\ninerm 1\kern-3.6pt\tenrm1}}%
\fi
\def\Im{{\rm\,Im\,}}    \def\Re{{\rm\,Re\,}}

\def\2{{1\over2}}

\def\igb{
    \mathop{\raise4.pt\hbox{\vrule height0.2pt depth0.2pt width6.pt}
    \kern0.3pt\kern-9pt\int}}

\def\st{\scriptscriptstyle}
\let\\=\noindent
\def\*{\vskip0.5truecm}
\def\?{\vskip0.75truecm}
\def\item#1{\vskip0.1truecm\parindent=0pt\par\setbox0=\hbox{#1\ }
     \hangindent\wd0\hangafter 1 #1 \parindent=\stdindent}

\def\annota#1#2{\footnote{${}^#1$}{\vtop 
{\hsize=\notesize\settepunti\baselineskip=8pt\parindent=0pt#2}}}
\def\annotano#1{\annota{\number\noteno}{#1}\advance\noteno by 1}
\newdimen\notesize\notesize=\hsize \advance \notesize by -\parindent


\def\ie{\hbox{\sl i.e.\ }}
\def\eg{\hbox{\sl e.g.\ }}
\def\qed{\hfill\break\nobreak\vbox{\vglue.25truecm\line{\hfill\raise1pt 
          \hbox{\vrule height9pt width5pt depth0pt}}}\vglue.25truecm}


\def\gint(#1)(#2)(#3){{\cal D}#1^{#2}\,e^{(#1^{#2+},#3#1^{#2-})}}

      \def\nn{{\bf n}}
  \def\V0{{\bf 0}}  \def\rr{{\bf r}}  \def\tt{{\bf t}}

\def\AA{{\cal A}}\def\BB{{\cal B}}
\def\DD{{\cal E}}\def\EE{{\cal E}}\def\FF{{\cal F}}
\def\GG{{\cal G}}\def\HH{{\cal H}}\def\II{{\cal I}}
\def\JJ{{\cal J}}\def\KK{{\cal K}}\def\LL{{\cal L}}
\def\MM{{\cal M}}\def\OO{{\cal O}}
\def\PP{{\cal P}}\def\QQ{{\cal Q}}\def\RR{{\cal R}}
\def\TT{{\cal T}}

\def\XX{{\cal X}}\def\YY{{\cal Y}}\def\hh{{f}}


\ifnum\cind=1
\def\prtindex#1{\immediate\write\indiceout{\string\parte{#1}{\the\pageno}}}
\def\capindex#1#2{\immediate\write\indiceout{\string\capitolo{#1}{#2}{\the\pageno}}}
\def\parindex#1#2{\immediate\write\indiceout{\string\paragrafo{#1}{#2}{\the\pageno}}}
\def\subindex#1#2{\immediate\write\indiceout{\string\sparagrafo{#1}{#2}{\the\pageno}}}
\def\appindex#1#2{\immediate\write\indiceout{\string\appendice{#1}{#2}{\the\pageno}}}
\def\paraindex#1#2{\immediate\write\indiceout{\string\paragrafoapp{#1}{#2}{\the\pageno}}}
\def\subaindex#1#2{\immediate\write\indiceout{\string\sparagrafoapp{#1}{#2}{\the\pageno}}}
\def\bibindex#1{\immediate\write\indiceout{\string\bibliografia{#1}{Bibliografia}{\the\pageno}}}
\def\preindex#1{\immediate\write\indiceout{\string\premessa{#1}{\the\pageno}}}
\else
\def\prtindex#1{}
\def\capindex#1#2{}
\def\parindex#1#2{}
\def\subindex#1#2{}
\def\appindex#1#2{}
\def\paraindex#1#2{}
\def\subaindex#1#2{}
\def\bibindex#1{}
\def\preindex#1{}
\fi

\def\leaderfill{\leaders\hbox to 1em{\hss . \hss} \hfill }


\newdimen\capsalto \capsalto=0pt
\newdimen\parsalto \parsalto=20pt
\newdimen\sparsalto \sparsalto=30pt
\newdimen\tratitoloepagina \tratitoloepagina=2\parsalto
\def\aboveparteskip{\bigskip \bigskip}
\def\belowparteskip{\medskip \medskip}
\def\abovecapitskip{\bigskip}
\def\belowcapitskip{\medskip}
\def\belowparskip{\smallskip}
%


\def\parte#1#2{
   \9{\immediate\write16
      {#1     pag.\numeropag{#2} }}
   \aboveparteskip 
   \noindent 
   {\ftitolo #1} 
   \hfill {\ftitolo \numeropag{#2}}\par
   \belowparteskip
   }


\def\premessa#1#2{
   \9{\immediate\write16
      {#1     pag.\numeropag{#2} }}
   \abovecapitskip 
   \noindent 
   {\it #1} 
   \hfill {\rm \numeropag{#2}}\par
   \belowcapitskip
   }


\def\bibliografia#1#2#3{
  \ifnum\stile=1
   \9{\immediate\write16
      {Bibliografia    pag.\numeropag{#3} }}
   \belowcapitskip
   \noindent 
   {\bf Bibliografia} 
   \hfill {\bf \numeropag{#3}}\par
  \else
    \paragrafo{#1}{References}{#3}
\fi
   }


\newdimen\newstrutboxheight
\newstrutboxheight=\baselineskip
\advance\newstrutboxheight by -\dp\strutbox
\newdimen\newstrutboxdepth
\newstrutboxdepth=\dp\strutbox
\newbox\newstrutbox
\setbox\newstrutbox = \hbox{\vrule 
   height \newstrutboxheight 
   width 0pt 
   depth \newstrutboxdepth 
   }
\def\newstrut
   {\relax \ifmmode \copy \newstrutbox \else \unhcopy \newstrutbox \fi}
%
%
\vfuzz=3.5pt
%
%
\newdimen\indexsize \indexsize=\hsize
\advance \indexsize by -\tratitoloepagina
\newdimen\dummy
\newbox\parnum
\newbox\parbody
\newbox\parpage
%

%

\def\mastercap#1#2#3#4#5{
   \9{\immediate\write16
      {Cap. #3:#4     pag.\numeropag{#5} }}
   \abovecapitskip
   \setbox\parnum=\hbox {\kern#1\newstrut{#2}
                        {\bf Capitolo~\number#3.}~}
   \dummy=\indexsize
   \advance\indexsize by -\wd\parnum
   \setbox\parbody=\vbox {
      \hsize = \indexsize \noindent \newstrut 
      {\bf #4}
      \newstrut \hss}
   \indexsize=\dummy
   \setbox\parnum=\vbox to \ht\parbody {
      \box\parnum
      \vfill 
      }
   \setbox\parpage = \hbox to \tratitoloepagina {
      \hss {\bf \numeropag{#5}}}
   \noindent \box\parnum\box\parbody\box\parpage\par
   \belowcapitskip
   }
\def\capitolo#1#2#3{\mastercap{\capsalto}{}{#1}{#2}{#3}}
%


%
\def\masterapp#1#2#3#4#5{
   \9{\immediate\write16
      {App. #3:#4     pag.\numeropag{#5} }}
   \abovecapitskip
   \setbox\parnum=\hbox {\kern#1\newstrut{#2}
                        {\bf Appendice~A\number#3:}~}
   \dummy=\indexsize
   \advance\indexsize by -\wd\parnum
   \setbox\parbody=\vbox {
      \hsize = \indexsize \noindent \newstrut 
      {\bf #4}
      \newstrut \hss}
   \indexsize=\dummy
   \setbox\parnum=\vbox to \ht\parbody {
      \box\parnum
      \vfill 
      }
   \setbox\parpage = \hbox to \tratitoloepagina {
      \hss {\bf \numeropag{#5}}}
   \noindent \box\parnum\box\parbody\box\parpage\par
   \belowcapitskip
   }
\def\appendice#1#2#3{\masterapp{\capsalto}{}{#1}{#2}{#3}}
%

%

\def\masterpar#1#2#3#4#5{
   \9{\immediate\write16
      {par. #3:#4     pag.\numeropag{#5} }}
   \setbox\parnum=\hbox {\kern#1\newstrut{#2}\number#3.~}
   \dummy=\indexsize
   \advance\indexsize by -\wd\parnum
   \setbox\parbody=\vbox {
      \hsize = \indexsize \noindent \newstrut 
      #4
      \newstrut \hss}
   \indexsize=\dummy
   \setbox\parnum=\vbox to \ht\parbody {
      \box\parnum
      \vfill 
      }
   \setbox\parpage = \hbox to \tratitoloepagina {
      \hss \numeropag{#5}}
   \noindent \box\parnum\box\parbody\box\parpage\par
   \belowparskip
   }
\def\paragrafo#1#2#3{\masterpar{\parsalto}{}{#1}{#2}{#3}}
\def\sparagrafo#1#2#3{\masterpar{\sparsalto}{}{#1}{#2}{#3}}
%


%
\def\masterpara#1#2#3#4#5{
   \9{\immediate\write16
      {par. #3:#4     pag.\numeropag{#5} }}
   \setbox\parnum=\hbox {\kern#1\newstrut{#2}A\number#3.~}
   \dummy=\indexsize
   \advance\indexsize by -\wd\parnum
   \setbox\parbody=\vbox {
      \hsize = \indexsize \noindent \newstrut 
      #4
      \newstrut \hss}
   \indexsize=\dummy
   \setbox\parnum=\vbox to \ht\parbody {
      \box\parnum
      \vfill 
      }
   \setbox\parpage = \hbox to \tratitoloepagina {
      \hss \numeropag{#5}}
   \noindent \box\parnum\box\parbody\box\parpage\par
   \belowparskip
   }
\def\paragrafoapp#1#2#3{\masterpara{\parsalto}{}{#1}{#2}{#3}}
\def\sparagrafoapp#1#2#3{\masterpara{\sparsalto}{}{#1}{#2}{#3}}
%


\ifnum\stile=1

\def\newcap#1{\setcap{#1}
\vskip2.truecm\advance\numsec by 1
\\{\ftitolo \numero\numsec. #1}
\capindex{\numero\numsec}{#1}
\vskip1.truecm\numfor=1\pgn=1\numpar=1\numteo=1\numlem=1
}

\def\newapp#1{\setcap{#1}
\vskip2.truecm\advance\numsec by 1
\\{\ftitolo A\numero\numsec. #1}
\appindex{A\numero\numsec}{#1}
\vskip1.truecm\numfor=1\pgn=1\numpar=1\numteo=1\numlem=1
}

\def\newpar#1{
\vskip1.truecm
\vbox{
\\{\bf \numero\numsec.\numero\numpar. #1}
\parindex{\numero\numsec.\numero\numpar}{#1}
\*{}}
\nobreak
\advance\numpar by 1
}

\def\newpara#1{
\vskip1.truecm
\vbox{
\\{\bf A\numero\numsec.\numero\numpar. #1}
\paraindex{\numero\numsec.\numero\numpar}{#1}
\*{}}
\nobreak
\advance\numpar by 1
}

\else

\def\newsec#1{\vskip1.truecm
\advance\numsec by 1
\vbox{
\\{\bf \numero\numsec. #1}
\parindex{\numero\numsec}{#1}
\*{}}\numfor=1\pgn=1\numpar=1\numteo=1\numlem=1
\nobreak
}

\def\newsubsect#1{
\vskip1.truecm
\vbox{
\\{\bf \numero\numsec.\numero\numpar. #1}
\parindex{\numero\numsec.\numero\numpar}{#1}
\*{}}
\nobreak
\advance\numpar by 1
}

\def\newapp#1{\vskip1.truecm
\advance\numsec by 1
\vbox{
\\{\bf A\numero\numsec. #1}
\appindex{A\numero\numsec}{#1}
\*{}}\numfor=1\pgn=1\numpar=1\numteo=1\numlem=1
}

\def\appendices{\numsec=0\def\Eq(##1){\Eqa(##1)}\def\eq(##1){\eqa(##1)}
\def\teq(##1){\teqa(##1)}}

\def\biblio{\vskip1.truecm
\vbox{
\\{\bf References.}\*{}
\bibindex{{}}}\nobreak\makebiblio
}

\fi


\newread\indicein
\newwrite\indiceout

\def\faindice{
\openin\indicein=\jobname.ind
\ifeof\indicein\relax\else{
\ifnum\stile=1
  \pagenumbersind
  \pageno=-1
  \setind
  \null
  \vskip 2.truecm
  \\{\ftitolo Indice}
  \vskip 1.truecm
  \parskip = 0pt
  \input \jobname.ind
  \ppaginan
\else
\\{\bf Index}
\*{}
 \input \jobname.ind
\fi}\fi
\closein\indicein
\def\nomeindice{\jobname.ind}
\immediate\openout \indiceout = \nomeindice
}


\newwrite\bib
\immediate\openout\bib=\jobname.bib
\global\newcount\bibex
\bibex=0
\def\verabib{\number\bibex}

\ifnum\tipobib=0
\def\cita#1{\expandafter\ifx\csname c#1\endcsname\relax
\hbox{$\clubsuit$}#1\write16{Manca #1 !}%
\else\csname c#1\endcsname\fi}
\def\rife#1#2#3{\immediate\write\bib{\string\raf{#2}{#3}{#1}}
\immediate\write15{\string\C(#1){[#2]}}
\setbox199=\hbox{#2}\ifnum\maxit < \wd199 \maxit=\wd199\fi}
\else
\def\cita#1{%
\expandafter\ifx\csname d#1\endcsname\relax%
\expandafter\ifx\csname c#1\endcsname\relax%
\hbox{$\clubsuit$}#1\write16{Manca #1 !}%
\else\probib(ref. numero )(#1)%
\csname c#1\endcsname%
\fi\else\csname d#1\endcsname\fi}%
\def\rife#1#2#3{\immediate\write15{\string\Cp(#1){%
\string\immediate\string\write\string\bib{\string\string\string\raf%
{\string\verabib}{#3}{#1}}%
\string\Cn(#1){[\string\verabib]}%
\string\CCc(#1)%
}}}%
\fi

\def\Cn(#1)#2{\expandafter\xdef\csname d#1\endcsname{#2}}
\def\CCc(#1){\csname d#1\endcsname}
\def\probib(#1)(#2){\global\advance\bibex+1%
\9{\immediate\write16{#1\verabib => #2}}%
}

\def\C(#1)#2{\SIA c,#1,{#2}}
\def\Cp(#1)#2{\SIAnx c,#1,{#2}}

\def\SIAnx #1,#2,#3 {\senondefinito{#1#2}%
\expandafter\def\csname#1#2\endcsname{#3}\else%
\write16{???? ma #1,#2 e' gia' stato definito !!!!}\fi}

\bibskip=10truept
\def\hboxto{\hbox to}

\catcode`\{=12\catcode`\}=12
\catcode`\<=1\catcode`\>=2
\immediate\write\bib<
        \string\halign{\string\hboxto \string\maxit%
        {\string #\string\hfill}&%
        \string\vtop{\string\parindent=0pt\string\advance\string\hsize%
        by -1.55truecm%
        \string#\string\vskip \bibskip
        }\string\cr%
>
\catcode`\{=1\catcode`\}=2
\catcode`\<=12\catcode`\>=12

\def\raf#1#2#3{\ifnum \bz=0 [#1]&#2 \cr\else
\llap{${}_{\rm #3}$}[#1]&#2\cr\fi}

\newread\bibin

\catcode`\{=12\catcode`\}=12
\catcode`\<=1\catcode`\>=2
\def\chiudibib<
\catcode`\{=12\catcode`\}=12
\catcode`\<=1\catcode`\>=2
\immediate\write\bib<}>
\catcode`\{=1\catcode`\}=2
\catcode`\<=12\catcode`\>=12
>
\catcode`\{=1\catcode`\}=2
\catcode`\<=12\catcode`\>=12

\def\makebiblio{
\ifnum\tipobib=0
\advance \maxit by 10pt
\else
\maxit=1.truecm
\fi
\chiudibib
\immediate \closeout\bib
\openin\bibin=\jobname.bib
\ifeof\bibin\relax\else
\raggedbottom
\input \jobname.bib
\fi
}

\openin13=#1.aux \ifeof13 \relax \else
\input #1.aux \closein13\fi
\openin14=\jobname.aux \ifeof14 \relax \else
\input \jobname.aux \closein14 \fi
\immediate\openout15=\jobname.aux

\def\V#1{\underline{#1}}

\def\normalbaselines{\baselineskip=20pt\lineskip=3pt\lineskiplimit=3pt}

\def\bi{{\bf i}}
\def\bj{{\bf j}}
\def\bk{{\bf k}}
\def\bl{{\bf l}}

\def\bo{{\bf o}}
\def\Id{{\rm Id}}
\def\rr{\right}
\def\ll{\left}


\titolo{Absolute
continuity of projected SRB measures of coupled Arnold cat map lattices}
\autore{F. Bonetto}{Department of Mathematics, Rutgers University,
New Brunswick, NJ 08903}
\autore{A. Kupiainen}{Department of Mathematics, 
Helsinki University, P.O  Box 4, Helsinki 00014, Finland}
\autore{J.L. Lebowitz}{Department of Mathematics and Physics,
Rutgers University, New Brunswick, NJ 08903}

\abstract{
We study a $d$-dimensional coupled map lattice consisting of   
hyperbolic toral automorphisms
(Arnold cat maps) that are weakly coupled by an analytic
coupling map. We construct the Sinai-Ruelle-Bowen measure
for this system and study 
its marginals on the  tori. We prove they are absolutely continuous
with respect to the Lebesgue measure if and only if
the coupling satisfies a nondegeneracy condition.}

\parole{dynamical systems, perturbation theory, projected measure,
absolute continuity}

\prima

\newsec{Introduction}

There has been much interest recently in time invariant measures of
physical systems evolving under certain types of non-Hamiltonian
deterministic dynamics.  These dynamics are chosen (invented) with the
intent of making these measures model the behavior of stationary
nonequilibrium states of real physical systems: \eg  the
``Gaussian thermostated'' dynamics \cita{EM}. An interesting example is
provided by the Moran and Hoover
model of electric current carrying systems\cita{MH}.  A
particle moves on a torus among fixed obstacles under the influence of
an external electric field $E$ and a {\it thermostat} which keeps the
energy fixed (it would otherwise grow indefinitely).  A very striking
(initially surprising) result of the numerical simulations was that
the stationary phase space density in a Poincare section looked very
``fractal'', i.e.\ singular with respect to the reference Lebesgue
measure. The singular nature of the invariant measure was later proven
rigorously, for $E \ne 0$, at least when $E$ is small  \cita{CELS}. Further
computer simulations and rigorous results (under suitable hypotheses)
strongly suggest that thermostated stationary measures are indeed
generically singular with respect to the Lebesgue measure\cita{SS}. They
correspond to the Sinai--Ruelle--Bowen (SRB) measures for these
systems \cita{SRB}.

A question then arose of what significance the singular nature of such
measures, so different from those obtained from the traditional
stochastic modeling of these systems, has for the behavior of
macroscopic nonequilibrium systems. While some authors attached
great significance to this fractality \cita{P}\cita{H} others
argued however that this is not the case\cita{WW}. The reason given
by the latter is that relevant
observable properties of macroscopic systems correspond to sums of
functions which depend only on the coordinates and velocities of one
or a few particles, \eg the electrical current is a sum over the
velocities of many interacting particles. Their steady state values
can therefore be computed from the reduced one or two particle
distribution functions and we expect these induced measures to be
absolutely continuous with respect to the Lebesgue measure. Of course to
make the thermostatted  dynamical systems appropriate for modeling
physical situations one would need to show that these reduced
distributions are equal, in the bulk, to those obtained from stochastic
boundary drives or from considering infinite system with
Hamiltonian dynamics. This is in fact what appears to be the case when
the Moran-Hoover model is extended to many particles\cita{BL}.

In this paper we prove the
absolute continuity of the reduced distributions or induced measure for 
a very idealized dynamical system made up of an infinite collection of
Arnold cat maps of the two torus, indexed by a $d$-dimensional
lattice. This dynamical system has typically an invariant
measure which is singular with respect to the Lebesgue measure.
We prove however that,
under general conditions, the projected measure on a single torus
is absolutely continuous with respect to Lebesgue
measure. Note that our result is for a projection on an explicitly
given surface on which the measure is singular in the absence of
coupling to other systems---not just for a ``typical'' projection.
This requires some condition on the interaction which we specify --
those excluded are very special and are essentially uncoupled
systems.

\newsec{Definitions and results}
 
The dynamical systems that we consider in this paper are so
called coupled map lattices \cita{PS}. 
The phase space of such a system is given in
general by a cartesian product over a $d$-dimensional lattice
$\Omega=\Zset^d$ of finite dimensional manifolds $\MM$. In our case,
$\MM$ is the two dimensional torus $\MM=\Tset=\Rset^2/\Zset^2$ and the
full phase space is $\TT=\Tset^{\Omega}$, equipped with the product
topology. We will construct the systems via finite dimensional
approximations, letting $\TT_N=\Tset^{\Omega_N}$ where
$\Omega_N=\Zset_N^d$ and $\Zset_N$ consists of integers of absolute
value strictly less than $N$.

The dynamics in a coupled map lattice is defined by first fixing a dynamical
system on each separate $\MM$ and then coupling them appropriately. In
the case at hand, let $A:\Tset\to \Tset$ be the Anosov dynamical system
defined by the linear transformation $A\in GL_2(\Zset)$ with $|{\rm det} A|=1$.  Letting $A$
act on each copy of $\Tset$ defines the {\it uncoupled} map $A:\TT\to
\TT$ and respectively on $\TT_N$.  Denoting by $\psi \in \Tset$ the
points on the two dimensional torus and by
$\Psi=(\Psi_\bi)_{\bi\in\Omega}$ those on $\TT$, the
Lebesgue measures $d\psi$ on each
torus and their product  $d\Psi$, are invariant for $A$.

To describe the coupled map, let $\hh:\TT\rightarrow \Rset^2$ be a
map and define $\AA:\TT\to\TT$ by
$$
\left(\AA\Psi\right)_\bi=A\Psi_\bi+\FF_{\bi}(\Psi)\qquad
\bi\in \Omega\Eq(nu)
$$
where 
$$
\FF_{\bi}(\Psi)=\hh(\tau_{-\bi}\Psi)\Eq(trans)
$$
and $\tau$ defines the $\Zset^d$-action on $\TT$ by
$(\tau_\bi\Psi)_\bj=\Psi_{\bi+\bj}$.  The pair $(\AA, \TT)$ defines
the coupled map lattice dynamical system.

To proceed we need to make assumptions on $\hh$.  We suppose the
coupling is weak and local, i.e. that $\FF_i$ depends weakly on
$\Psi_\bj$ for $\bj$ far away from $\bi$. A convenient way to encode
this is to assume $\hh$ is holomorphic with derivatives with respect
to $\Psi_\bi$
decaying rapidly with $\bi$. Given two positive constants
$\alpha$ and $\beta$, let $\Tset_{\bi,\a,\beta}
\subset\Cset^2/\Zset^2$ be the complex neighbourhood of $\Tset$ defined by
$|\Im\Psi_\bi|<\a e^{|\bi|\beta}$, and $\RR$ the
cartesian product of the $\Tset_{\bi,\a,\beta}$. 
If $H$ is the space of holomorphic
functions $\hh:\RR\to\Cset$ equipped with the norm

$$\Vert\hh\Vert_\infty=\sup_{\Psi\in\RR}|\hh(\Psi)|.\Eq(norm)$$
we will consider the dynamical system eqs. \equ(nu),\equ(trans) 
with $\hh\in H$ of $\Vert\hh\Vert_\infty$ sufficiently small. 

This infinite dimensional dynamical system will be studied via
finite dimensional approximations which we now define.
Letting $\RR_N$ be the cartesian product of the $\Tset_{\bi\a,\beta}$
for $|\bi|<N$ and 
given an $\hh\in H$ we let $\hh_N$ be the map defined on $\RR_N$ given by
 
$$\hh_N(\Psi)=\hh(\Psi^p)$$
where $\Psi^p\in\RR$ is obtained
by extending $\Psi\in\RR_N$ periodically to $\RR$ .  We define the finite
dimensional approximation $\AA_N$ to $\AA$ by equations
\equ(nu)\equ(trans) where $\tau$ is the action of translations modulo
$(N\Zset)^d$, \ie we impose periodic boundary conditions on
$\Omega_N$. Observe that $\AA$ maps  the set 
$\PP_N\subset\RR$ of periodic points of period $N$ to itself. Thus,
identifying $\RR_N$ with $\PP_N$ we have 

$$\AA_N\equiv \AA\large|_{\PP_N}.\Eq(AN)$$
We define the SRB measure for $\AA_N$ ($\AA$ respectively) to be the weak limit
of $\AA_N^n m_N$ ($\AA^n m$) as $n\to\infty$ of the normalized Lebesgue measure
$m_N$ ($m$) on $\TT_N$ ($\TT$) if such a limit exists.  
Our first result concerns the existence of a SRB measure for $\AA$.

\*
\0{\bf Theorem 1:\ }{\it There exists an $\e>0$ such that given
$\hh\in H$ with $\Vert\hh\Vert\le \e$ the dynamical systems $\AA_N$
have a SRB measure $\mu_N$ for each $N\le\i$. The weak limit of
$\mu_N$ as $N\to\infty$ exists and is equal to $\mu$. The
measures $\mu_N$ and $\mu $ are $C^\i$ smooth in $\hh$ in the ball
$\Vert\hh\Vert<\e$ of $H$ in the sense that $\int T d\mu$ is
$C^\i$ smooth for any $C^\i$ smooth $T$ depending on finitely many
variables $\Psi_\bi$.  }
\*

\noindent{\bf Remark.} The existence of the $N\to\infty$ limit
of the SRB measures has been proven before \cita{JP}, with lesser
regularity assumptions than here. However, we need more detailed
structure of the measures and have to go through the construction.

\bigskip

Let ${\bf P}$ be the projection of $\TT_N$ to the torus
at origin $\Tset$  and ${\bf P}\mu_N$ the induced 
projection of $\mu_N$ on $\Tset$. We want to
address the question whether this projection is absolutely continuous
with respect to the Lebesgue measure on $\Tset$.

\bigskip

\noindent {\bf Definition}. {\it $\AA_N$ is {\rm degenerate}
if for all $\Psi\in \TT_N$ the unstable manifold of $\Psi$ is a
cartesian product of curves $\gamma_\bi(\Psi)$ lying on the $\bi^{\rm th}$
torus.}

\bigskip

An example of a degenerate map is the uncoupled map: in
this case the curve $\gamma_\bi(\Psi,\xi)=\Psi_\bi+e^+\xi$ for
$\xi\in\Rset$ where $Ae^+=\Lambda_+e^+$ with $\Lambda_+>1$. More
generally if we choose $f(\Psi)=g(\Psi)e^+$ with $g:\TT\rightarrow
\Rset$ it is easy to see that the map $\AA$ given by
eqs.\equ(nu),\equ(trans) with such an $f$ has the same unstable
foliation as $A$. In this case we will say that $\AA$ is {\it coupled
through the unstable manifold}.  We can characterize all
degenerate coupled maps through the following

\*
\0{\bf Proposition 1.  }{\it  $\AA_N$ is degenerate if and only if there
exists $X:\Tset\rightarrow\Tset$ such that $X\circ\AA_N\circ
X^{-1}=\tilde \AA_N$ where $\tilde \AA_N$ is coupled
through the unstable manifold.}
\*

Our main result is

\*
\0{\bf Theorem 2.  }{\it For each $2\le N\le\infty$ if $\AA_N$ is not degenerate then the
projected measures ${\bf P}^*\mu_N$ are absolutely continuous with respect to
the Lebesgue measure on $\Tset$.
Moreover if $\AA$ is degenerate then $\AA_N$ is degenerate for
every $N$ and if $\AA$ is nondegenerate then $\AA_N$ is too for $N$ 
large enough.}
\*

We close this section with a remark concerning the fractality of
$\mu_N$.  The Hausdorff dimension of $\mu_{N}$ will generically
satisfy ${\rm dim}_{HD}\mu_{N} <{\rm dim}\TT_N$.  In fact from
the Kaplan-Yorke formula \cita{KYY} one obtains the upper bound

$${\rm dim}_{HD}\mu_{N}\le {\rm dim}\TT_N+{\mu_{N}(\sigma)\over
\l_{min}}\Eq(HDN)$$
where $\l_{min}$ is the minimum Lyapunov exponent of $\AA_{N}$ and
$\sigma(\Psi) = -\log(\det D\AA_N(\Psi))$. Generically we expect that 
${\mu_{N}(\sigma)/\l_{min}}\ge\delta\dim\TT_N$ for some constant
$\delta$. Indeed it is easy to show that for a generic perturbation of $A$
acting on $\Tset$ $\mu_1(\sigma)>0$, see
\cita{BGM}. Adding a small enough coupling we will have
$\mu_N(\sigma)\simeq N\mu_1(\sigma)$ while $\l_{min}$ is almost
independent from $N$. Theorem 2 asserts then that, notwithstanding this
extensive loss of dimensionality of the attractor, the projected SRB
measure is still absolutely continuous. In particular this argument
shows that our theorem is not empty when $N=\infty$.

\newsec{The conjugacy}

We start by constructing a conjugacy $X: \TT\to\TT$ of the coupled map
$\AA$ to the uncoupled one $A$:

$$X\circ A=\AA\circ X\Eq(conj)$$
Observe that, form \equ(AN) it follows that $X_N\equiv X\big|_{\PP_N}$
conjugates $\AA_N$ to $A_N$.

Given a map $x:\TT\to\Rset^2$ let $\displaystyle\tau
x:\TT\to(\Rset^2)^{\sZset^d}$ be defined by translations as $(\tau
x)_\bi=x\circ\tau_{-\bi}$. With this notation, $\FF=\tau\hh$. Hence,
guided by translation invariance of our map $\AA$ we look for a
solution of eq.\equ(conj) in the form $X={\rm Id}+\tau x$ with
$x$ solution of the equation
$$
{\bf T}x=\hh({\rm Id}+\tau x)\Eq(conj')
$$
where ${\bf T}$ is the linear operator defined by
$$
{\bf T}x=x\circ A-A\circ x . \Eq(T)
$$
We expect from general theory that the solution $x$ will not
be a differentiable function of $\Psi$ but only H\"older continuous.
Given a function $g:\TT_N\to\Rset^2$ let $\delta_\bj$ denote the
``H\"older derivative''
$$
\delta_\bj
g(\Psi)=\sup_{v_\bj}{\left| g(\Psi+v_\bj)-
g(\Psi)\right|\over |v_\bj|^{\g}}
$$
where $\gamma<1$ and the supremum runs over vectors having a nonzero
component only at the $\bj^{\rm th}$ position and of length no larger
than unity. Form now on we fix $\g<1$ and, to avoid cumbersome
notation, do not indicate the dependence of the estimates in what follows on
$\gamma$ as well as on $\alpha$ and $\beta$. Moreover
we will use $C$ to indicate the constants that appear in all the estimates.

Let $\EE$ be the Banach space of H\"older continuous maps
$x:\TT\to\Rset^2$ with norm

$$\Vert x\Vert=\Vert x\Vert_\infty+\sum_{\bj} e^{{_\beta\over^2}|\bj|}
\Vert \delta_\bj x\Vert_\infty.\Eq(C^a)$$
We then have
\*
\0{\bf Proposition 2:\ }{\it There exists an $\e>0$ such that
given $\hh\in H$ with $\Vert\hh\Vert \le \e$ equation \equ(conj')
has a unique solution in $\EE$ with $\Vert x\Vert\le C\Vert\hh\Vert$.
Moreover $x$ is analytic in $\hh$ in the ball $\Vert\hh\Vert <\e$.}
\*

\0{\bf Proof.} Let us call ${\bf H}x={\bf T}^{-1}\hh({\rm Id}+\tau x)$. 
We want to show that ${\bf H}$ is a contraction in the ball $B=\{x|
\Vert x\Vert\le R\Vert\hh\Vert\}$ for a suitable $R$.

It is easy to find an explicit representation for ${\bf
T}^{-1}$. Let $e^+,\Lambda^+$ and $e^-,\Lambda^-$ denote the two
eigenvectors of the matrix $A$ and the corresponding eigenvalues, with
$\Lambda^+>1$ and $\Lambda^-={{\rm det }A\over\Lambda^+}$, where $|{\rm det }A|=1$.  $e^+$ and $e^-$ are
the unit vectors in the direction of the unstable and stable
manifolds at each point $\psi \in \Tset^2$.  Expressing vectors
$v\in\Rset^2$ in this basis as $v=v_+e^++v_-e^-$, we have

$$\ll({\bf T}^{-1}x\rr)(\Psi)=\sum_{n=0}^{\i}\Lambda_-^n
x_+\ll(A^{-n+1}\Psi\rr)+\sum_{n=1}^{\i}\Lambda_+^{-n}
x_-\ll(A^{n-1}\Psi\rr)\Eq(exp)$$  
From this expression it follows immediately that the norm of ${\bf
T}^{-1}$ as an operator in $\EE$ is bounded by

$$\Vert{\bf T}^{-1}\Vert_{L(\EE,\EE)}\le
{4\over 1-\Lambda_+^{-(1-\g)}}\Eq(inverse)$$
We now claim that the function $h_x(\Psi)= \hh(\Psi+\tau x(\Psi))$
satisfies

$$\Vert h_x\Vert\le C\Vert\hh\Vert,\qquad
\Vert h_x-h_y\Vert\le C\Vert\hh\Vert\,\Vert x-y\Vert.\Eq(claims)$$
To prove the first inequality in \equ(claims) we write
$$
|h_{x}(\Psi+v_\bj)-h_{x}(\Psi)|= 
\sum_{\bk}\int_0^1 dt\,\partial_\bk 
\hh\ll(\Psi^t\rr)\ll(v_{\bj,\bk}+
x\ll(\tau_{-\bk}(\Psi+v_\bj)\rr)-x(\tau_{-\bk}\Psi)\rr)
$$
where $\Psi^t=\Psi+tv_\bj+ t\tau x(\Psi+v_\bj)+(1-t) \tau x(\Psi)$ and
$v_{\bj,\bk}$ is the $\bk$ component of $v_\bj$.
Then, using  $|\partial_\bk \hh|\le
e^{-\beta |\bk|}\Vert\hh\Vert$, which follows from \equ(norm), and
$$
|x\ll(\tau_{-\bk}(\Psi+v_\bj)\rr)-
x(\tau_{-\bk}\Psi)|\le \eta^\gamma  \Vert\delta_{\bj-\bk}x\Vert_\infty,\Eq(st1)
$$
where we set $\eta=|v_\bj|$, we get
$$
\sum_\bj e^{{\beta\over^2}|\bj|}\eta^{-\g}|h_{x}(\Psi+v_\bj)-h_{x}(\Psi)|
\le \Vert\hh\Vert
(\sum_{\bj} e^{-{_\beta\over^2}
|\bj|} +
\sum_{\bj\bk} e^{{_\beta\over^2}|\bj|}
e^{-\beta |\bk|}\Vert\delta_{\bj-\bk}x\Vert_\infty).\Eq(aaa)
$$
>From \equ(C^a) we infer $\Vert\delta_{\bj-\bk}x\Vert_\infty\le
e^{-{_\beta\over^2}|\bk-\bj|}\Vert x\Vert$.
Hence by a use of the triangle inequality \equ(aaa) is bounded by
$$
C\Vert\hh\Vert +\Vert\hh\Vert 
\Vert x\Vert\sum_\bk e^{-{_\beta\over^2}|\bk|}
\le C\Vert\hh\Vert(1+\Vert x\Vert).\Eq(stima)
$$

The second inequality of \equ(claims) can be proven as follows. Observe that,

$$ 
h_x(\Psi)-h_y(\Psi)=\int_0^1 dt \partial_\bk f(\Psi+\tau
x(\Psi)+(1-t) \tau y(\Psi))(x(\tau_\bk\Psi)-y(\tau_\bk\Psi))\Eq(st2)
$$
so that

$$\Vert h_x-h_y \Vert  \le \sum_\bk \Vert \partial_\bk 
f(Id+\tau x+(1-t) \tau y)\Vert
 \Vert x-y \Vert \Eq(stima1) $$
Combining eq.\equ(norm) with a Cauchy estimate we infer for $\Psi$
real
$$
|\partial_\bk\partial_\bi f(\Psi)|\le e^{-\beta(|\bk|+|\bj|)} 
\Vert f \Vert_\infty\Eq(cau2)
$$ 
Proceeding as above this implies that
$$
\Vert \partial_\bk f(Id+\tau x+(1-t) \tau y)\Vert\le Ce^{-\beta|\bk|}
\Vert f \Vert_\infty\Eq(st3)
$$  
and \equ(claims) follows.
Eqs. \equ(inverse) and \equ(claims) establish the contractive property
for suitable $R$.  By the Banach fixed point theorem we have a unique
solution of eq.\equ(conj') which is analytic in $\hh$.

\newsec{The invariant manifolds}

In this section we will construct the two invariant manifolds
$W^\pm(\Psi)$ defined, for every point $\Psi\in \TT$, by the property
$$
W^\pm(\Psi)=\left\{\Psi'| \lim_{n\rightarrow \infty}
|\AA^{\mp n}\Psi-\AA^{\mp
n}\Psi'|=0\right\}.\Eq(invama)
$$
where $|\Psi|=\sup_\bi |\Psi_\bi|$. We observe again that the stable
and unstable manifolds of $\AA_N$ are given by the periodic points in
$W^\pm(\Psi)$. We will give below a unified construction of these sets
for $N\le \infty$, $N=\infty$ referring to $W^\pm(\Psi)$. For
convenience the $N$-dependence of the various objects will be
suppressed whenever possible.

We shall look for $W^\pm(\Psi)$  in terms of an embedding
$$
\xi\in\Rset^{\Omega_N}\rightarrow S_\Psi^\pm(\xi)\in 
\ll(\Rset^2\rr)^{\Omega_N} \Eq(S)
$$
($\Omega_\infty$ means $\Zset^d$) such that the action of $\AA$ is given by
$$
\AA S_\Psi^\pm(\xi)=S_{\AA\Psi}^\pm(\tilde\LL^\pm(\Psi)\xi)
\Eq(inva)
$$
where $\tilde\LL^\pm(\Psi)$\ are linear operators on
$\Rset^{\Omega_N}$. For $N=\infty$ we mean by the latter the vector
space $\ell_\infty(\Zset^d)$ .

We want to use eq.\equ(inva) to study the regularity properties of
$S^\pm$ as a function of $\Psi,\xi$ and $\AA$.  We expect on general
grounds $S^{\pm}$ to be at most $C^{\a}$ in $\Psi$.  Thus, since $\AA$
occurs in eq.\equ(inva) coupled to $\Psi$, low regularity can be
expected for $S$ also as a function of $\AA$.  However, it will be
convenient to have maximal regularity in $\AA$ and this can be
achieved by looking for the solution to \equ(inva) in the form
$$
S_\Psi^\pm(\xi)=\Psi+\XX^\pm(X^{-1}(\Psi),\xi)\Eq(S1)
$$
where $X$ is the conjugation constructed in Section 3.
Eq. \equ(inva) implies the following equation for $\XX^{\pm}$:
$$
\AA(X(\Psi)+\XX^\pm(\Psi,\xi))=X(A\Psi)+
\XX^\pm(A \Psi,\LL^\pm(\Psi)\xi)\Eq(inva1)
$$
where $\LL=\tilde\LL\circ X$ and the previous problem is clearly not
present. Indeed, we will show that \equ(inva1) has a solution
$\XX^{\pm}$ that is {\it analytic} in $\AA$ and in $\xi$ as well.

To state the main result of this section we need to introduce the
space where \equ(inva1) will be solved. Let $D_N$ be the complex domain
$D_N=\{\xi||\xi_\bi|<1, \forall \bi\in\Omega_N\}$.  Let $\BB$ be the
Banach space of maps $\XX: \TT_N\times D_N\rightarrow
(\Cset^2)^{\Omega_N}$ which are H\"older continuous 
in $\Psi$ and analytic in $\xi$
equipped with the norm 

$$\Vert \XX\Vert=\sup_{\bi} (\Vert
\XX_{\bi}\Vert_\infty+\sum_{\bj}
e^{{_\beta\over^4}|\bi-\bj|}\Vert D_\bj\XX_{\bf
\bi}\Vert_\infty ).\Eq(bnorm)$$

where $D_\bj=(\delta_{\bj}, \partial_{\xi_\bj})$ and the infinity
norm is intended in both $\Psi$ and $\xi$.  The following Proposition
describes the local stable and unstable manifolds:
 
\*
\0{\bf Proposition 3:\ }{\it There exists an $\e>0$, independent of 
$N\le\infty$ such that given $\hh\in H$ with $\Vert\hh\Vert \le \e$
the local stable and unstable manifolds $W^\pm(\Psi)$ are given by
real analytic embeddings
$$
S_\Psi^\pm : D_N\rightarrow \left(\Rset^2\right)^{\Omega_N}. 
$$
$S^\pm_\Psi$ are translation invariant: $S^\pm_{\tau_i\Psi}(\tau_i\xi)=
S^\pm_\Psi(\xi)$ and are given by \equ(S1) with $\XX^\pm
\in\BB$ and

$$\Vert \XX^\pm-\Lambda_+\xi\Vert \le C\Vert\hh\Vert.$$
Moreover $\XX^\pm$ are analytic functions of $\hh$ in the ball
$\Vert\hh\Vert <\e$ of the Banach space $H$. }
\*

To describe the global result let $\MM$ be the
Banach space of $C^\alpha$ maps $\LL$ from $\TT_N$ to the linear
operators on $\Rset^{\Omega_N}$ equipped with the norm

$$\Vert \LL\Vert=\sup_{\bi}\left(
\sum_\bj e^{{\b\over 4}|\bi-\bj|}\Vert \LL_{\bi\bj}
\Vert_\infty+\sum_{\bj\bk} e^{{_\beta\over^4}|\bi-\bk|}
\Vert\delta_\bk\LL_{\bi\bj}\Vert_\infty \right).\Eq(C^b)$$
We have then
\*
\0{\bf Proposition 4:\ }{\it With the assumptions of Proposition 2,
the global stable and unstable manifolds $W^\pm(\Psi)$ are given as
real analytic embeddings 
$$
S_\Psi^\pm :  \Rset^{\Omega_N}\rightarrow (\Rset^2)^{\Omega_N} 
$$
that satisfy equation \equ(inva1) with $\LL\in\MM$ and
$$
\Vert\LL^\pm-\Lambda_\pm\Vert\le C\Vert\hh\Vert\Eq(stL)
$$
Moreover $S_\Psi^\pm$ can be extended to a complex neighborhood of
$(\Rset^2)^{\Omega_N}$.}
\*
 
\0{\bf Proof:\ } We start the proof by separating the linear part in 
$\xi$ from the rest in $\XX^\pm(\Psi,\xi)$, \ie we write

$$\XX^\pm(\Psi,\xi)=\chi^{\pm}(\Psi)\xi+
\bar\XX^{\pm}(\Psi,\xi)\Eq(svi)$$
Observe that $\chi^{\pm}(\Psi)$ is a linear map from
$\Rset^{\Omega_N}$ to ${\rm
T}_\Psi\TT_N$. We will choose as a basis on ${\rm
T}_\Psi\TT_N$ the one formed by the vectors $e^-_{\bf i}$ and
$e^+_{\bf i}$.

The matrix $\chi^{\pm}(\Psi)$ satisfies the equation:

$$
\AA\chi^{\pm}(\Psi)-\chi^{\pm}(A \Psi)\LL^\pm=
D\FF({X(\Psi)})\chi^{\pm}(\Psi)\Eq(lin)
$$
>From now on we will consider explicitly only the unstable ($+$) case
and drop the $+$ superscript.  Identical considerations hold for the
stable manifold.  It is easy to see that eq.\equ(lin) alone cannot
fix uniquely $\chi$ and $\LL$ . In fact if the pair $\chi(
\Psi),\LL(\Psi)$ is a solution of eq.\equ(lin) then,
given any nonvanishing function $l:\TT_N\to\Rset$, 
$$
\chi'(\Psi)=l(\Psi)\chi(\Psi)\qquad\LL'(\Psi)=
{l(\AA \Psi)\over l(\Psi)} \LL(\Psi)\Eq(ambi)
$$
is also a solution. To resolve the above ambiguity we fix $\chi_{+}={\rm
Id}$ where the subscript $+$ refers to the component along the
unstable directions and, with a slight abuse, we denote the $-$
component $\chi_{-}$ by $\chi$. Thus $\chi$ is now a $\Omega_N\times
\Omega_N$ matrix. Writing the matrix $H(\Psi)=D\FF({X(\Psi)})$ in 
the $\pm$ basis so that
$$
D\AA=\pmatrix{\Lambda_+\Id +H_{++}&H_{+-}\cr H_{-+}
&\Lambda_+^{-1}\Id +H_{--}}\Eq(decom)
$$
it follows that
$$
\eqalignno{\Lambda_+ +H_{++}+H_{+-} \chi(\Psi)-
\LL(\Psi)=0&\eq(++)\cr
H_{-+}+(\Lambda_+^{-1}+H_{--}) \chi(\Psi)-
\chi(A \Psi)\LL(\Psi)=0&\eq(--)
}
$$
Setting now 
$$
\LL(\Psi)=\Lambda_+{\rm Id}+\bar\LL(\Psi)\Eq(barL)
$$ 
we may solve eq.\equ(++) for $\bar\LL(\Psi)$:
$$ 
\bar\LL(\Psi)=H_{++}+H_{+-} \chi(\Psi)\Eq(sol1)
$$
and substituting this in eq.\equ(--) we get
$$
{\bf T}_1\chi(\Psi)=H_{--} \chi(\Psi)+
H_{-+}-\chi (A \Psi) H_{++}-
\chi(A \Psi) H_{++}
\chi (\Psi)\equiv F(\chi,\Psi)\Eq(+++)
$$
where ${\bf T}_1$ is the operator
$$
({\bf T}_{1}\chi)(\Psi)=\Lambda_+ \chi(A \Psi)-
\Lambda_+^{-1}\chi(\Psi).\Eq(T1)
$$
We solve this equation in the Banach space $\MM$ with the norm
eq. \equ(C^b). The inverse of ${\bf T}_1$ is given by
$$
{\bf T}_1^{-1} \chi(\Psi)=\sum_{n=0}^{\infty}\Lambda_+^{-2n-1} \chi
(A^{-n-1}\Psi)\Eq(TT)
$$
from which follows that ${\bf T}_1$ is a bounded operator in $\MM$.
Note that due to the extra power of $\Lambda_+^{-1}$ compared to the
eq. \equ(exp) we could work in $C^1$. This gain is not useful because
$F(\chi,\Psi)\in C^\a$. 

The solution of \equ(+++) proceeds analogously to what was done in the
previous section. Writing it as $\chi={\bf T}_1^{-1}F(\chi)$ we show
the right hand side is contraction in $\Vert\chi\Vert\le
C\epsilon_0$. This follows in a straightforward fashion using the
following Lemmas.
\*
\0{\bf Lemma 1:\ }{\it $\MM$ is a Banach algebra:
$$ 
\Vert\chi\eta\Vert\le 2\Vert\chi\Vert\Vert\eta\Vert\Eq(alge)
$$ 
}
\*
\0{\bf Proof}. The claim follows from the simple estimates

$$\sum_{\bj} e^{{\b\over 2}|\bi-\bj|}|
(\chi\eta)_{\bi\bj}|\le 
\sum_{\bj\bl} e^{{\b\over 2}(|\bi-\bl|+|\bl-\bj|)}
|\chi_{\bi\bl}||\eta_{\bl\bj}|\le \Vert\chi\Vert\Vert\eta\Vert\Eq(al1)
$$
and in a similar manner
$$
\sum_{\bj\bk} e^{{\b\over 2}|\bi-\bk|}
|\partial_\bk(\chi\eta)_{\bi\bj}|\le 
\sum_{\bj\bk\bl} \ll(e^{{_\beta\over^2}|\bi-\bk|)}
|\partial_\bk\chi_{\bi\bl}||\eta_{\bl\bj}|+
e^{{_\beta\over^2}|\bi-\bl|}
|\chi_{\bi\bl}|e^{{_\beta\over^2}|\bl-\bk|}
|\partial_\bk\eta_{\bl\bj}|\rr)\le 
2\Vert\chi\Vert\Vert\eta\Vert.\Eq(al2)
$$
\*
\*
\0{\bf Lemma 2:\ }{\it For $i,j=\pm$ we have $H_{i,j}\in \MM$.}
\*
\0{\bf Proof}: note first that from eq.\equ(norm) we get
$$
|\partial_\bk\FF_\bi(\Psi)|\le C\Vert\hh\Vert
e^{-\beta|\bi-\bk|}\Eq(b1)
$$
and 
$$
|\partial_\bl\partial_\bk\FF_\bi(\Psi)|\le C\Vert\hh\Vert
e^{-\beta(|\bi-\bk|+|\bi-\bl|)}\Eq(b2)
$$
for $\Psi\in \RR$. It follows that
$$
|\delta_\bk\HH_{\bi,\bj}(\Psi)|=\left|\sum_\bl\partial_\bj\partial_\bl
\FF_\bi\big|_{X(\Psi)}\delta_\bk X_\bl(\Psi)\right|
\le C\a^{-1}\Vert\hh\Vert^2
e^{-{\beta\over 2} |\bi-\bk|}.\Eq(b3)
$$
These are summable when multiplied by the exponential factors
in our norm.
\*
\*
To summarize 
\*
\0{\bf Proposition 5:\ }{\it There exists an $\e$ such that given
$\hh:\RR\to\Rset^{2\Omega_N}$ with $\Vert\hh\Vert \le \e$
equation \equ(lin) has a unique solution $\chi=(1,\chi_-)$ with
$\chi_-\in\MM$ and $\Vert
\chi_-\Vert \le C\Vert\hh\Vert$. $\LL$ is given
by \equ(barL) with $\Vert
\bar\LL\Vert \le C\Vert\hh\Vert$.  Moreover $\chi$ and $\LL$ are
analytic in $\hh$ in the ball $\Vert\hh\Vert <\e$.}
\*

Let us finally consider the remainder $\bar\XX$ in eq. \equ(svi).
Using eq. \equ(lin) 
we deduce
$$
\bar\XX(A \Psi,\LL(\Psi)\xi)-A\bar\XX(\Psi,\xi)=G(\Psi,\xi,\bar\XX)
\Eq(rima)
$$
where
$$
G(\Psi,\xi,\bar\XX)\equiv\FF\left(X(\Psi)+
\chi(\Psi)\xi+\bar\XX(\Psi,\xi)\right)-\FF(X(\Psi))-
D\FF(X(\Psi))\chi(\Psi)\xi\Eq(defG)
$$
Let ${\bf T}_2$ be the operator
$$
{\bf T}_2\bar\XX(\Psi,\xi)=\bar\XX(A \Psi,\LL(\Psi)\xi)-
A\bar\XX(\Psi,\xi).\Eq(T2)
$$
Thus we need to solve the equation
$$
\bar\XX={\bf T}_2^{-1}G(\bar\XX)\Eq(Geq)
$$
in the Banach space $\BB$ with norm given by \equ(bnorm). 
First we need to control the inverse of ${\bf T}_2$ given formally by
$$
({\bf T}_2^{-1}\bar\XX)(\Psi,\xi)=\sum_{n=0}^{\infty}A^n
\bar\XX(A^{-n-1}\Psi,\widehat\LL^{n}(\Psi)\xi)\Eq(inve)
$$
where $\widehat\LL^{n}(\Psi)=\prod_{i=1}^{n-1}\LL(A^{-i}\Psi)$. 
Recall that $\bar\XX$ vanishes at $\xi=0$ together with its first
derivatives, i.e. we want to solve our equation in the closed subspace
$\BB_0$ of $\BB$ of functions with this property.
We first prove
\*
\0{\bf Lemma 3:\ }{\it The map
$$
{\bf F}:\bar\XX\to A \bar\XX(A^{-1}\Psi,\LL(\Psi)^{-1}\xi)\Eq(lem3)
$$
is a bounded map from $\BB_0$ into itself with norm strictly less than
one.}
\*
\0{\bf Proof.} From $\LL=\Lambda_++\bar\LL$ and
$\Vert\bar\LL\Vert\le C\Vert f\Vert$ we infer that if $\xi\in D_N$
then $\LL(\Psi)^{-1}\xi\in \rho D_N$ with 
$$
\rho=(\Lambda_+-C\Vert
f\Vert)^{-1}.
$$
 Hence by a Cauchy estimate, taking into account that
$\bar\XX(\Psi,\xi)$ vanish to second order for $\xi=0$, we get 
$$
\Vert{\bf
F}\bar\XX_i\Vert_\i\le\l\Vert\bar\XX_i\Vert_\i 
$$ 
for $\l=\Lambda_+\rho^2<1$
provided $\Vert f\Vert$ is chosen small enough.

For the second factor occuring in the norm \equ(bnorm) we write
$$
\eqalign{&\sum_{\bj}e^{{\beta\over 4}|\bi-\bj|}
\Vert D_\bj\bar\XX_\bi(A^{-1}\Psi,\LL(\Psi)^{-1}\xi)\Vert_\infty\le
\rho^2\Lambda_+^\g\sum_{\bj}e^{{\beta\over 4}|\bi-\bj|}\Vert\delta_{\bj}
\bar\XX_\bi(\Psi,\xi)\Vert_\infty+\cr
+&\rho\sum_{\bk,\bl,\bj}e^{{\beta\over 4}|\bk-\bj|}\Vert\delta_{\bj}
\ll(\LL(\Psi)^{-1}\rr)_{\bk,\bl}\Vert_\infty e^{{\beta\over 4}|\bk-\bi|}
\Vert\partial_{\xi_\bk}\bar\XX_\bi(\Psi,\xi)\Vert_\infty+\cr
+&\rho\sum_{\bj,\bk}e^{{\beta\over 4}|\bk-\bj|}
\Vert\ll(\LL(\Psi)^{-1}\rr)_{\bk,\bj}\Vert_\infty
e^{{\beta\over 4}|\bi-\bk|}\Vert\partial_{\xi_\bk}\bar\XX_\bi(\Psi,\xi)\Vert_\infty
}
$$
where the factors $\rho^2$ and $\rho$ come from a Cauchy estimate on
$\rho D_N$. By the definitions of the norms \equ(bnorm) and \equ(C^b)
the sums may be bounded by
$$
(\rho^2\Lambda_+^\g+\rho\Vert \LL(\Psi)^{-1}\Vert)\Vert\bar\XX\Vert
$$
and since 
$$
\Vert \LL(\Psi)^{-1}\Vert\le (\Lambda_+-C\Vert
f\Vert)^{-1}
$$ 
the claim follows with $\Vert f \Vert$ small enough.
\*\*
Hence ${\bf T}_2$ has a bounded inverse in $\BB_0$ as long as $\g<1$.

Next we turn to the study of $\Vert G\Vert$.
Note that $G$ is well defined: the argument of $\FF$ in
\equ(rima) is in its analyticity domain if $C\Vert f\Vert < \alpha$.
Moreover we want to prove that:
$$
\Vert G(\bar\XX)\Vert\le C\Vert f\Vert \Vert\bar\XX \Vert\qquad\qquad 
\Vert G(\bar\XX)-G(\bar\YY)\Vert\le C\Vert f\Vert\Vert\bar\XX -\bar\YY\Vert
\Eq(claim2)
$$
so that we can conclude our prove invoking again the Banach fixed
point theorem.

To prove the above estimates we must bound both the derivatives in
$\xi$ and the H\"older derivative in $\Psi$ of $\GG$. It is easy to
see that the $\xi$ derivatives bound follows easily from Cauchy type
estimates like eqs.\equ(b1)\equ(b2).  To bound the H\"older derivative
in $\Psi$ we observe that for both of the above estimates it is enough
to study the first term in the definition \equ(defG) since good
estimates were already proven on the other two terms while proving
the existence of $X$ and $\chi$. To this end, using
$\HH(\Psi,\bar\XX)=\HH\left(X(\Psi)+
\chi(\Psi)\xi+\bar\XX(\Psi,\xi)\right)$, we can write:
$$
\eqalign{\HH_\bi&(\Psi,\bar\XX)-\HH_\bi(\Psi+\delta v_\bi,\bar\XX)=
\sum_{\bk} \int_0^1 dt\,\partial_\bk 
\HH_\bi\ll(\Psi^t\rr)\cdot\cr&
(v_{\bj,\bk}+(X(\Psi+v_\bj)-X(\Psi))+
(\chi(\Psi+v_\bj)\xi-\chi(\Psi)\xi)+
(\bar\XX(\Psi+v_\bj,\xi)-\bar\XX(\Psi,\xi)))}\Eq(st21)
$$
where we have set 
$$
\Psi^t=tv_{\bj,\bk}+t(X(\Psi+v_\bj)+\chi(\Psi+v_\bj)\xi+
\bar\XX(\Psi+v_\bj,\xi))+(1-t)(X(\Psi)+\chi(\Psi)\xi+
\bar\XX(\Psi,\xi))\Eq(st22)
$$
and proceed like in eq. \equ(stima). The second inequality follows
from
$$\eqalign{
\HH(\Psi,\xi,\bar\XX)-\HH(\Psi,\xi,\bar\YY)=&
\int_0^1 dt \partial_\bk \FF(\Psi+X(\Psi)+\chi(\Psi)\xi+
t\bar\XX(\psi)\cr &+(1-t) \bar\YY(\Psi))(\bar\XX(\Psi)-\bar\YY(\Psi))}\Eq(st23)
$$
and again we can conclude like in eq.\equ(stima1).

For Proposition 3 note that the analyticity domain of $\XX^+$ in $\xi$
is independent of $\Psi$. Eq.\equ(inva1) implies
$$
\XX(\Psi,\xi)=\AA (X(A^{-1}\Psi)+\XX(A^{-1}
\Psi,\LL(A^{-1}\Psi)^{-1}\xi))-X(\Psi)\Eq(pro)
$$
so that by Lemma 3 the right hand side provides analytic continuation 
of the left hand side
to $\rho D_N$ with $\rho=(\Lambda_+-C\Vert f\Vert)$. Iterating this
formula $n$ times we expand the domain of $\XX^+$ as long as
$X(A^{-1}\Psi)+\XX^\pm(A^{-1}\Psi,\LL^+(A^{-1}\Psi)^{-1}\xi)$ is in
the analyticity domain of $\AA$. Since $\AA=A+\FF$ the imaginary part
of $\XX$ may expand each step by a factor $\Lambda_++C\e_0$. Hence for
$\Re \xi \in \rho_n D_N$ with $\rho_n=(\Lambda_+-C_1\e_0)^n$, we can take
$\Im \xi \in r_n D_N$ with
$r_n=(\Lambda_++C_2\e_0)^{-n}$. Thus $\XX^+$ is analytic in $\xi$
in such a neighborhood of $\Rset^{\Omega_N}$. Furthermore, since
$W^\pm_\FF(\Psi)=X_\FF(W^\pm_0(\Psi))$, as follows immediately from the
definition of the unstable manifold, the continuity of $X$ and density
of $W^+_0(\Psi)$ imply that $W^+_\HH(\Psi)$ is dense in $\TT_N$.

\newsec{The SRB measure}

The SRB measure is constructed in a standard way using a Markov
partition. Since we want to have a construction uniform in $N$ and
also keep track of analyticity properties in that limit we can not
refer directly to standard constructions. However, we assume
the reader is familiar with the various standard definitions
concerning Markov partitions and thermodynamic formalism and will use
them freely without comment\cita{Ru}.

Let $Q=\{Q_i\}_{1,\ldots,m}$ be a Markov partition of the two-torus
$\Tset$ corresponding to the linear map $A$.  We recall that the $Q_i$
are standard rectangles in $\Rset^2$ with sides parallel to the
vectors $e^\pm$.

Let $S_N=\{1,\ldots,m\}^{\Omega_N}$. Then ${\bf Q}=\{{\bf Q}_s\}_{s\in
S_N}$ where ${\bf Q}_s=\times_{\bi\in\Omega_N}Q_{s(\bi)}$ is a Markov
partition for $A$ acting on $\TT_N$ and
$$
\QQ=\{{\QQ}_s\}_{s\in
S_N}\qquad {\QQ}_s=X({\bf Q}_s)\Eq(mar)
$$
is a Markov partition for $\AA$.

As usual, a Markov partition allows to conjugate $\AA$ to a subshift
of finite type on a symbol sequence space.  Let
$\bar\Sigma_N=S_N^{\sZset}$ and denote its elements by
$\sigma=\{\sigma_i\}_{i\in \sZset}$ where $\sigma_i\in S_N$ is written
as $\sigma_i=(\sigma_i(\bj))_{\bj\in
\Omega_N}$. The fact that $\QQ$ is a Markov partition implies
that the set
 $$
\PP(\sigma)=\cap_{i\in\sZset}\AA^{-i}
(\QQ_{\sigma_i})\Eq(sym)
$$
contains at most one point. Let $\Sigma_N$ be the set of all $\sigma$
such that $\PP(\sigma)$ contains exactly one point (we will call this
point $\PP(\sigma)$ with a small abuse of notation). The Markov
property of $\QQ$ and the way we constructed it imply that there exist
a $m\times m$ matrix $M$ with $M_{ij}\in\{0,1\}$ such that
$\sigma\in\Sigma_N$ if and only if
$M_{\sigma_i(\bj),\sigma_{i+1}(\bj)}=1$ for every $i\in\Zset$ and
$\bj\in\Omega_N$. If we equip $\Sigma_N$ with the metric
$$
d(\sigma,\sigma')=\sum_{i,\bj}2^{-(|i|+|\bj|)}|\sigma_i(\bj)-\sigma'_i(\bj)|.
\Eq(dists)
$$
Then we have 

\*
\0{\bf Proposition 6:\ }{\it The map $\PP:\Sigma_N\to\TT_N$
is given by $\PP_\bi=p\circ\tau_{-\bi}$ where $p: \Sigma_N\to\Tset$ and

$$|p(\sigma)-p(\sigma')|\le Cd(\sigma,\sigma')^\eta$$

for a suitable H\"older exponent $\eta$.
Moreover $\PP$ conjugates $\AA$ to the shift $\tilde\tau$ on $\Sigma_N$, \ie $(\tilde \tau \sigma)_i=\sigma_{i-1}$.
}
\*
\noindent{\bf Proof.} Let $\PP_0(\sigma)$ be the map associated with
$A$. It is clear that $\PP_0(\sigma)=p_0\circ\tau_{-\bi}$ and that
$p_0$ depends only on the value of $\sigma$ at the origin $\bo$ of
$\Zset^d$. For this map the time part of the estimate is a simple
consequence of the hyperbolicity of $A$. Our theorem follows
immediately from the fact that $\PP(\sigma)=X(\PP_0(\sigma))$ and the
H\"older continuity of $X$ proved in section 3.
\*\*
Observe that if we consider the metric on $\TT$ given by:

$$
d(\Psi,\Psi')=\sum_{\bj}2^{-|\bj|}|\Psi_\bj-\Psi'_\bj|.
\Eq(dist)
$$
then $\PP$ is an H\"older function from $\Sigma$ to $\TT$.

The SRB measure is constructed in the standard fashion by studying the
Jacobean of the map $\AA$ restricted to the unstable foliation. Recall
that the local unstable manifold at $\Psi$ is given by the embedding
\equ(S). We will use as a basis of the tangent space $TW^+(\Psi)$ the
vectors $\partial_{\xi_\bj}$, $\bj\in\Omega_N$. In this basis the
Jacobean of $\AA$ restricted to the unstable foliation is given at the
point $\Psi$ by $\det \tilde\LL(\Psi)$. Thus, let us define 

$$
\lambda^+(\Psi)=-\log\det(\Lambda_+^{-1}\LL(X^{-1}(\Psi)))\Eq(lam)
$$
where the constant $\Lambda_+^{-1}$ was inserted for later
convenience, and let
$$
h^+(\sigma)=\lambda^+(\PP(\sigma))
$$

Then we have

\*
\0{\bf Proposition 7:\ }{\it $\lambda^+$ and $h^+$
can be written as a sum of local functions as follows:
$$
\lambda^+(\Psi)=\sum_{\bi\in \Omega_N} \lambda(\tau_{\bi}\Psi)\Eq(Laml)
$$
and 
$$
h^+(\sigma)=\sum_{\bi\in \Omega_N} h(\tau_{\bi}\sigma)\Eq(hl)
$$
with $\lambda$ and $h$ H\"older continuous with constants
uniform in $N$. Furthermore
$$
|\lambda(\Psi)-\lambda(\Psi')|\le C\Vert\hh\Vert d(\Psi,\Psi')^\eta
\Eq(hold)$$ 
and
$$
|h(\sigma)-h(\sigma')|\le C\Vert\hh\Vert d(\sigma,\sigma')^\eta
\Eq(hold1)
$$
}
\*

\noindent{\bf Proof.} Writing
$$
\lambda^+(\Psi)={\rm Tr}\log\ll(1+\bar\LL(X(\Psi))\Lambda_+^{-1}\rr)=
{\rm Tr}\sum_{i=1}^\i {(-1)^i\over i}{\bar\LL(X(\Psi))^i\over \Lambda_+^i}
$$

we can define

$$
\lambda(\Psi)=\sum_{i=1}^\i{(-1)^i\over i}
{\ll(\bar\LL\ll(X(\Psi)\rr)^i\rr)_{\bo,\bo}
\over \Lambda_+^N}
$$
From Lemma 1 and Proposition 3 we get $\Vert\lambda(\Psi)\Vert_\i<C
\Vert f \Vert$ and $\Vert \delta_\bi
\lambda(\Psi)\Vert_\i<Ce^{-{\beta\over 4}|\bi|}$ from which
eq.\equ(hold) follows immediately. Eq.\equ(hold1) is an immediate
consequence of \equ(hold) and Proposition 5.
\*\*
The SRB measure of our system will be given in terms of a Gibbs state
on $\Sigma_N$. Let $e$ be the maximum entropy measure on
$\Sigma_N$ and define the ``Hamiltonian''
$$
H_T(\sigma)=\sum_{i=-T}^T h^+(\tilde\tau^i(\sigma))\Eq(Ham)$$ 
Set 
$$
\mu^T(d\sigma)={1\over Z_T}e^{H_T(\sigma)}e(d\sigma)\Eq(LT)
$$
where $Z_T=\int e^{H_T}de$
\*
\0{\bf Proposition 8:} {\it 
The weak limits
$$
\lim_{n\to\infty}
\AA_N^n m_N=\tilde\mu_N \qquad \lim_{T\to\infty}\mu^T
=\mu_N
$$
exist and $\tilde\mu_N=\PP\mu_N$. Furthermore, 
$\mu_N$ and $\tilde\mu_N$ converge weakly to measures $\mu$
and $\tilde\mu$ as $N\to\i$. }  
\*

\noindent{\bf Proof.} For any finite $N$ the maps $\l^+$
and $h^+$ are H\"older continuous. For instance
$$
|\l^+(\Psi)-\l^+(\Psi')|\le \sum_\bj |\l(\tau_\bj\Psi)-\l(\tau_\bj\Psi')|
\le C(N)\sup_\bj d(\tau_\bj\Psi,\tau_\bj\Psi')^\eta
$$
and the last distance is bounded by $C(N)d(\Psi,\Psi')$
as is readily seen from \equ(dist). The Bowen-Ruelle theorem
\cita{Bo} yields the claim for $\AA_N^n m_N$.

The claim for $\mu^T$ can be proven similarly, but let us prove
a more general result that comprises both the $T$ and the $N$
limits. Consider the Hamiltonian
$$
H_{T,N}(\sigma)= \sum_{i=-T}^T\sum_{\bj\in \Omega_N}
h(\tau_\bj\tilde\tau^i(\sigma)).\Eq(HTN)
$$
Given a $\sigma\in\Sigma_N$
let $\sigma^{n}\in \Sigma_N$ be defined
as $\sigma^{n}_i(\bj)=\sigma_i(\bj)$ for $|\bj|\le n$ and
$\sigma^{n}_i(\bj)=\sigma_i(0)$ for $|\bj|> n$. Write
$$
h(\sigma)=h(\s^0)+\sum_{n=1}^{n(N)} (h(\sigma^n)-h(\sigma^{n-1}))
\equiv\sum_{n=0}^{n(N)} h_n(\sigma)\Eq(loc) 
$$
and then do a similar telescoping sum in the time
direction\annotano{Some care should be paid here to take into account
the compatibility matrix $M$. This is a standard construction, see \eg
\cita{Ga}.}\ for each $h^n(\sigma)$ arriving at
$$
h(\sigma)=\sum_R h_R(\sigma)\Eq(squares)
$$
where $R$ are sets of the from $\{(i,\bj)|
|i|\le m, |\bj|\le n\}\in \Zset\times\Omega_N$ and 
$h_R$ depends on $\sigma$ only through its restriction
to $R$. The H\"older continuity expressed by eq. \equ(hold) of $h$
implies
$$
|h_R|\le C\Vert\hh\Vert e^{-cd(R)}\Eq(exp1)
$$
where $d(R)$ is the diameter of $R$. For the full Hamiltonian
we get now
$$
H_{T,N}(\sigma)=\sum_R h_R(\sigma)\Eq(local)
$$
where  $R$ are rectangles similar to the ones appearing in
eq.\equ(squares) but centered arbitrarily
in $[-T,T]\times \Omega_N$. For the existence of the limit
$$
\lim_{N\to\i}\lim_{T\to\i}e^{H_{T,N}}e\Eq(lim)
$$ 
(in any order, indeed) we refer the reader to e.g.  \cita{BK}
where it is proven in our setup provided that $\Vert\hh\Vert$ is small 
enough. We should warn the reader that
standard high temperature expansion methods can not be used
when the interactions have a decay as in eq. \equ(exp1)
where only the diameter of the set $R$ occurs (one needs the
volume of $R$). See \cita{BK} for a discussion of these subtleties.

Finally we have to prove that $\lim_{N\to\i}\tilde \mu_N=\tilde
\mu$. To do this one can use the symbolic map $\PP$. Some care should
be paid to the fact that $\PP$ is not one to one. Indeed the points on
the set
$$
\partial_\i\QQ=\bigcup_{n=-\i}^{\i}\bigcup_{s}\partial\QQ_s
$$
have more than one symbolic representation. Hence we need to show
that for every $s$ and $N$ we have $\mu_N(\partial\QQ_s)=0$. For
$N<\i$ this is evident while for $N=\i$ this follows easily with an
argument similar to that used to prove point (b) of proposition 11 below.

\newsec{Decomposition of the SRB measure}

\newsubsect{Coordinates on rectangles}

In order to study the projection of the SRB measure on finitely many
tori we need to express it in terms of our parameterization of the
stable and unstable manifolds constructed in sec.4.  To do this we
will introduce new coordinates on the rectangles $\QQ_{s}$.  For
$\Psi\in\QQ_s$ let

$$W^\pm_s(\Psi)=W^\pm(\Psi)\cap\QQ_s.\Eq(Ws)$$

Let us fix an arbitrary point $\psi_i$ on each basic rectangle
$Q_i$ of the 2-torus. Observe that $Q_i=U_i\times S_i$ where $U_i$ and
$S_i$ are segments in the direction of $e^+$ and $e^-$, respectively,
containing $\psi_i$. We set
$\bar\Psi_{s}=(\psi_{s(\bj)})_{\bj\in\Omega}\in{\bf Q}_{s}$ and
call $\Psi_{s}=X(\bar\Psi_{s})$ the {\it center} of $\QQ_{s}$. From
the fact that $\QQ_{s}$ is a rectangle we know that for every
$\Psi\in\QQ_{s}$ there is one and only one $\Psi'\in W^-_s(\Psi_s)$
such that $\Psi\in W^+_s(\Psi')$.  Hence there exists a unique
$\xi^-\in \Rset^{\Omega_N}$ such that $\Psi'=S^-_{\Psi_{s}}(\xi^-)$
and a unique $\xi^+\in \Rset^{\Omega_N}$ such that
$\Psi=S^+_{\Psi'}(\xi^+)$.  Thus we have a one to one map $\Psi\in
\QQ_{s}\rightarrow
(\xi^-,\xi^+)\in\Rset^{\Omega_N}\times\Rset^{\Omega_N}$ whose inverse
we will, with slight abuse, denote by $\Psi^N(s,\xi^-,\xi^+)$ i.e.
$$
\Psi^N(s,\xi^-,\xi^+)=S^+_{S^-_{\Psi_{s}}(\xi^-)}(\xi^+)
\Eq(coor)
$$
$\Psi^N$ can be viewed as a continuous map $\MM_N\to\TT_N$
where $\MM_N$ is a compact subset of $S_N\times\Rset^{\Omega_N}
\times\Rset^{\Omega_N}$ given by
$$
\MM_N=\{(s,\xi^-,\xi^+)|s\in S_N, \xi^-\in I_N(s), \xi^+\in J_N(s,\xi^-)\}\Eq(sp)
$$
where
$$
I_N(s)=(S^-_{\Psi_s})^{-1}W^-_s(\Psi_s)\Eq(I)
$$
and
$$
J_N(s,\xi^-)=(S^+_{\Psi'})^{-1}W^+_s(\Psi'),\qquad \Psi'=
S^-_{\Psi_{s}}(\xi^-).\Eq(J)
$$
Denoting the points in $\MM_N$ by $m$, we have by translation invariance
(see Proposition 3)
$$
\Psi^N_{\bi}(m)=\Psi^N_{0}(\tau_{-\bi}m).
$$
It is easy to see from the properties of the maps $S^\pm_\Psi$
that there exists an $r$ independent on $N$ s.t. $\MM_N
\subset S_N\times C^N_r\times C^N_r=\widehat{\MM}_N$ where $C^N_r$ is
the cube of side $r$ centered at origin of $\Rset^{\Omega_N}$.

Setting $C^\i_r$ equal to the $r$-cube in $\Rset^\Omega$ and giving it
the topology defined by the metric
$$
d(\xi,\xi')=\sum_\bj 2^{-|\bj|}|\xi_\bj-\xi'_\bj|.\Eq(disxi)
$$
and $S_\i=\{1,\ldots,m\}^\Omega$ with the metric
$$
d(s,s')=\sum_\bj 2^{-|\bj|}|s(\bj)-s'(\bj)|.\Eq(distsym)
$$
we have that $C^\i_r$ and $S_\i$ are  compact metric spaces. We can
view $\MM$ as a compact subset of $\widehat{\MM}$.

The following Lemma summarizes the important properties of the function $\Psi^N$.
\*
\0{\bf Proposition 9.} {\it There exist an $r$ such that $\Psi^N$
can be extended to a function from $\widehat{\MM}_N$ to $\TT$, still
denoted with $\Psi^N$. 
For every $s\in S_N$, $\Psi^N(s,\xi^-,\xi^+)$ is one to one from
$C^N_r\times C^N_r$ into its image. Moreover $\Psi^N_{0}$ converge as $N\to\i$ 
uniformly to a  H\"older continuous function $\Psi_0$. Finally,
for each $(s,\xi^-)$, $\Psi_0(s,\xi^-,\xi^+)$ is analytic
in $\xi^+$ for $|{\rm Im }\xi_\bi|<1$.}  
\*

\noindent{\bf Proof}: The extensions follows from the fact that
$\xi^+$ and $\xi^-$ are global coordinates on the unstable and stable
manifold. Moreover the image of $C^N_r\times C^N_r$ under $\Psi^N$ is
close to ${\bf Q}_s$ for every $s$ from which the one to one property
follows.  

The regularity property in $\xi^\pm$ immediately
follows from prop. 1 while the regularity in $s$ is a consequence of
the construction of the center $\Psi_s$, see definition after
eq. \equ(Ws). 
\*\*

Let us spell out the correspondence between the coordinates
$(\xi^-,s,\xi^+)$ and the symbolic  representation. Define
$$
C_{s}=\{\sigma|\sigma_0=s\}.\Eq(iden)
$$
On $C_{s}$ we have coordinates $\sigma^\pm\in S_N^{\sZset^\pm}\equiv
\Sigma_N^\pm$
where ${\Zset^\pm}$ are the strictly positive (negative) integers and
$$
(\sigma^-,\sigma^+)\to \sigma^-\vee s\vee\sigma^+
$$
is one to one $\Sigma_N^-\times\Sigma_N^+\to C_s$. Clearly
 $\PP(C_{s})=\QQ_s$ and given a point
$\bar\sigma\in C_s$ with $\PP(\bar\sigma)=\Psi$ then 
$$\eqalign{
W^+_s(\Psi)=&\{\PP(\bar\sigma^-,s,\sigma^+)|\sigma^+\in
\Sigma_N^+\}\cr 
W^-_s(\Psi)=&\{\PP(\sigma^-,s,\bar\sigma^+)|\sigma^-\in
\Sigma_N^-\}\cr}\Eq(map)
$$

Now the map $\Theta_N$ given by
$$
\Theta_N(\sigma)=\Psi_N^{-1}(\PP(\sigma))
$$
gives the desired correspondence between the two coordinate systems.

\newsubsect{Decomposition in finite volume}

Our goal is to find a
representation of the SRB measure in terms of the coordinates
$(\xi^-,s,\xi^+)$. Let us define 
$$
\nu_N=\Theta_N\mu_N 
$$
We will decompose the measure $\mu_N$ in a convolution of different
probability measures and then discuss their image under the map $\Theta_N$. 
Since the volume
$N$ is kept fixed in this subsection, we will omit it in the notation.
Recall that $\mu=\lim_{T\to\infty}\mu^T$ with
$$
\mu^T(d\sigma)={1\over Z_T}e^{H_T(\sigma)}e(d\sigma).
$$
Write $\sigma=\s^-\vee s \vee \sigma^+$ and decompose the maximum
entropy measure as
$$
e(d\sigma)=e(d\sigma^-|s)e(d\sigma^+|s)b(ds)\Eq(decome)
$$
where  $e(d\s^\pm|s)$ are the measure $e$ on $\Sigma_N^\pm$
conditioned on $ s$ and $b$ is the Bernoulli measure on $S_N$.
Similarly decompose the Hamiltonian 
$$
H_T(\s)=H_T^+(\s)+H_T^-(\s)
$$
into terms depending mostly on the $\s^+$ or  $\s^-$:
$$
H_{T}^+(\s)=\sum_{i=1}^T h^+(\tau^i\s)\qquad 
H_{T}^-(\s)=\sum_{i=0}^T h^+(\tau^{-i}\s).\Eq(hampm)
$$
Define on $\Sigma_N^+$ the probability measure, depending parametrically
on $s$ and $\s^-$:
$$
\mu^T_{s,\sigma^-}(d\sigma^+)={1\over
Z_{T}(s,\sigma^-)}e^{H^+(\s^-\vee s \vee \sigma^+)}e(d\sigma^+|s)\Eq(mu^p)
$$
where
$$
Z^{T}(s,\sigma^-)=\int e^{H^+(\s^-\vee s \vee
\sigma^+)}e(d\sigma^+|s).\Eq(normu^p)
$$
Let $\sigma_s=\s_s^-\vee s \vee \sigma^+_s$ be the symbolic representation
of the center $\Psi_s$ of $\QQ_s$ and set
$$
I_T(\s)=H_{T}^-(\s)-H_{T}^-(\s^-\vee s \vee \sigma^+_s).
$$
We can then write our measure as
$$
\mu^{T}(d\sigma)=e^{I_T(\s)}\mu^T_{s,\sigma^-}(d\sigma^+)
\mu^T_{s}(d\sigma^-)b(ds)\Eq(pm)
$$
where
$$
\mu^T_{s}(d\sigma^-)={Z_{T}(s,\sigma^-)\over Z_T}
e^{H_{T}^-(\s^-\vee s \vee \sigma^+_s)}e(d\s^-|s).\Eq(mu^-)
$$

The following Proposition characterizes the images
under $\Theta$ of $\mu^T_{s,\sigma^-}$, $\mu^T_{s}$ and
$I_{T}$.

\*
\0{\bf Proposition 10\ } {(a)\it The limit $\mu_{s,\s_-}=
\lim_{T\to\infty}\mu^T_{s,\s_-}$
exists, and $\nu_{s,\xi^-}=\Theta
\mu_{s,\sigma^-}$ is the normalized Lebesgue measure 
$|J_N(s,\xi^-)|^{-1}d\xi^+$ on $J_N(s,\xi^-)$ where
$\xi^-$ is given by 
$\Theta(\sigma^-\vee s\vee\sigma_s^+)=(\xi^-,s,0)$.}  
\*
\noindent (b)
{\it The limit $\mu_{s}=
\lim_{T\to\infty}\mu^T_{s}$
exists, and $\nu_{s}=\Theta
\mu_{s}$ is a positive Borel measure of finite mass on $I_N(s)$.}  
\*
\noindent (c) {\it The functions $I_T\circ \Theta^{-1}$
converge uniformly on $\MM_N$ to a H\"older continuous function
$\II(s,\xi^-,\xi^+)$. The function $\II(s,\xi^-,\xi^+)$ can be
extended to a H\"older continuous function $\widehat{\MM}_N$}  
\*

\noindent{\bf Proof.} Since these claims are rather standard
we will be brief.

\noindent (a) Let 

$$
\bar H_{T}(\s)=\sum_{i=1}^T h^+(\tau^i\s)+\sum_{i=0}^T
h^-(\tau^{-i}\s)
$$
where $h^-(\tau^{-i}\s)=\lambda^-(\PP(\sigma))$ with
$$
\lambda^-(\Psi)=-\log\det(\Lambda_-^{-1}\LL^-(X^{-1}(\Psi)))\Eq(lamr)
$$
Define the measure
$$
\bar\mu^T(d\sigma)={1\over \bar Z_T}e^{\bar H_{T}(\s)}e(d\sigma)
$$ 
and its image
$\bar\nu^T=\Theta_N\bar\mu^T$. It is well known that
$\bar\nu\=\lim_{T\to\infty}\bar\nu^T$ exists and is absolutely continuous with
respect to the Lebesgue measure with a continuous density. Thus, its
restriction to $\QQ_s$ is given in the $\xi^\pm$ coordinates as
$$\bar\nu_s(d\xi^+,d\xi^-)=g_s(\xi^+,\xi^-)d\xi^+d\xi^-$$
with $g$ continuous.

On the other hand we may decompose $\bar\nu$  as we did above $\mu$ and
get
$$
\bar\nu_s(d\xi^+,d\xi^-)=e^{\bar \II(s,\xi^-,\xi^+)}\nu_{s,\xi^-}(d\xi^+)
\bar\nu_s(d\xi^-)
$$
for some Borel measure $\bar\nu_s$ and continuous $\bar\II$. 
Hence we conclude that
$\nu_{s,\xi^-}$ is absolutely continuous with respect to the
Lebesgue measure on $J_N(s,\xi^-)$:
$$
\nu_{s,\xi^-}(d\xi^+)=f_{s,\xi^-}(\xi^+)d\xi^+ 
$$
where 
$f_{s,\xi^-}(\xi^+)$
is continuous in all variables. 

Let now $\AA_u$ be the map $\AA$
restricted to the unstable manifold. We get then 
$$(\AA_u\nu_{s,\xi^-})(d\xi^+_1)=
\det(\tilde\LL^+(\Psi(s,\xi^-,\xi^+)))^{-1}f_{s,\xi^-}(\xi^+)
d\xi^+_1 ,\Eq(cam)
$$
where $\AA(\Psi(s,\xi^-,\xi^+))=\Psi(s_1,\xi^-_1,\xi^+_1)$.

On the other hand, from definition \equ(mu^p) one concludes
$\tau\mu_{s,\sigma^-}=z(s,\sigma^-)
e^{h^+}\mu_{s_1,\sigma^{-}_1}$ where $\tau\sigma=
\sigma_1$ and
 $z(s,\sigma^-)=\lim_{T\to\i}Z_{T-1}(s,\sigma^-)Z_T^{-1}(s,\sigma^-)$.
Thus
$$(\AA_u\nu_{s,\xi^-})(d\xi^+_1)=\tilde z(s,\xi^-)
\det(\tilde\LL^+(\Psi(s,\xi^-,\xi^+)))^{-1}f_{s,\xi^-}(\xi^+_1)
d\xi^+_1 \Eq(cam1)
$$
and therefore $f_{s_1,\xi^-_1}(\xi^+_1)=\tilde z(s,\xi^-)f_{s,\xi^-}(\xi^+)$. 
Fixing now $s,\xi^-$ let $s_{-n},\xi^-_{-n}$ and $J_{-n}\subset 
J(s_{-n},\xi^-_{-n})$ be such that $\AA^n_u$ maps $J_{-n}$ bijectively
onto $J(s,\xi^-)$. Then
$$f_{s,\xi^-}(\xi^+)=\prod_{i=1}^n\tilde z(s_{-i},\xi^-_{-i})
f_{s_{-n},\xi^-_{-n}}(\xi^+_{-n})
$$
with $\xi^+_{-n}\in J_{-n}$. By expansiveness of $\AA_u$
the intervals $J_{-n}$ shrink exponentially and the RHS
converges to a $\xi^+$ independent limit, which then is
fixed by the fact that $\nu_{s,\xi^-}$ is a probability
measure.

\*
\noindent (b) 
In statistical mechanics terms $\mu_s$ is the Gibbs measure for spins
$\sigma^-$ in the half space of negative time, with $s\vee\sigma^+$
as boundary conditions in nonnegative times. The $T\to\i$ limit then
follows from exponential decay of interactions guaranteed by the
H\"older property of $h^+$. 
\*
\noindent (c)
We have 
$$
\II(s,\xi^-,\xi^+)=\lim_{T\to\i}\sum_{i=0}^T
\l^+(\AA^{-i}(\Psi^N(s,\xi^-,\xi^+)))-\l^+
(\AA^{-i}(\Psi^N(s,\xi^-,0)))\Eq(gg)
$$
By H\"older continuity of $\l^+$ the summand is bounded in absolute value by
$$
C(N)d(\AA^{-i}(\Psi^N(s,\xi^-,\xi^+)),\AA^{-i}(\Psi^N(s,\xi^-,0)))
\le C(N)2^{-i\eta}.
$$
Hence the limit as $T\to\i$ exists. The extension follows immediately
form the representation eq.\equ(gg).
\*\*

To summarize, the SRB measure in the $m$ coordinates is given as
$$
\nu(dm)=e^{\II(m)}1_{J(s,\xi^-)}(\xi^+)
b(ds)\nu_s(d\xi^-)
{d\xi^+\over |J(s,\xi^-)|}.\Eq(decomf)
$$

\newsubsect{Decomposition in the infinite volume limit.}

We are interested in the limit as $N\to\i$ of the above measures but to
study the projected SRB measure we will decompose $\nu_s$ extracting
from it a finite dimensional part $\xi_M$ of the unstable coordinate
$\xi^+$. Thus let us fix an integer $M$ and for $N>M$ write
$\Rset^{\Omega_N}=\Rset^{\Omega_M}\times
\Rset^{\Omega_N\setminus\Omega_M}$ and $\xi^+=(\xi_M,\xi^\perp)$
accordingly. The actual value of $M$ we need to study the projected
SRB measure will be fixed in the following section. We can rewrite
eq.\equ(gg) as 
$$
\eqalign{
\II_N(m)=&\lim_{T\to\i}\sum_{i=0}^T\ll(
\l^+(\AA^{-i}(\Psi^N(s,\xi^-,\xi^+)))-\l^+
(\AA^{-i}(\Psi^N(s,\xi^-,(0,\xi^\perp)))\rr)+\cr
+&\lim_{T\to\i}\sum_{i=0}^T\ll(
\l^+(\AA^{-i}(\Psi^N(s,\xi^-,(0,\xi^\perp)))-\l^+
(\AA^{-i}(\Psi^N(s,\xi^-,0)))\rr)\cr
\equiv & \JJ_N(m)+\KK_N(m')}.\Eq(gm)
$$
where the triple $(s,\xi^-,\xi^\perp)$ was denoted by $m'$. Let
$\MM'_N$ be the set of all $m'=(s,\xi^¯,\xi^\perp)$ such that
$(s,\xi^¯,\xi^\perp,\xi_M)\in\MM_N$ for some $\xi_M$. Clearly 
$\MM'_N\subset S_N\times C_r^{N,M}=\widehat{\MM}'_N$ where
$C_r^{N,M}$ is the cube of side $r$ in $\Rset^{\O_N/\O_M}$.
Given $\xi^\perp\in C_r^{N,M}$, we set 
$$
\{\xi_M|(\xi_M,\xi^\perp)\in J_N(s,\xi^-)\}
\equiv J_N(m')\subset  C_r^M\Eq(para)
$$
while given $\xi_M\in C_r^{M}$, we set
$$
\{\xi^\perp|(\xi_M,\xi^\perp)\in J_N(s,\xi^-)\}
\equiv J^\perp_N(s,\xi^-,\xi_M)\subset  C_r^{N,M}\Eq(perp)
$$
Finally let the projection of the set $J_N(s,\xi^-)$ to 
the $\xi^\perp$ direction, i.e. to $\Rset^{\Omega_N\setminus\Omega_M}$
be denoted by $J^\perp_N(s,\xi^-)$. Clearly we have
$$
J^\perp_N(s,\xi^-)=\bigcup_{\xi_M\in C_r^M}
J^\perp_N(s,\xi^-,\xi_M).
$$

We may then rewrite the SRB measure \equ(decomf) as
$$
\nu_N(dm)=\rho_N(dm')\vartheta^N_{m'}(d\xi_M)
,\Eq(decomf1)
$$
where
$$\eqalign{
\rho_N(dm')=&e^{\KK_N(m')}1_{J_N^\perp(s,\xi^-)}(\xi^\perp)b(ds)\nu_{Ns}(d\xi^-)
{d\xi^\perp\over |J_N(s,\xi^-)|}\cr
\vartheta^N_{m'}(d\xi_M)=&e^{\JJ_N(m)}1_{J_N(m')}(\xi_M)d\xi_M.}
$$
Clearly for every finite $N$ and every continuous
function $T_N$ on $\MM_N$ we can write 
$$
\int T_N(m)\nu_N(dm)=\int \rho_N(dm')\int \vartheta^N_{m'}(d\xi_M)T_N(m)
$$
We now want to show that we can take the limit of this identity.
\*
\0{\bf Proposition 11\ }:{\it There exist a bounded H\"older continuous function $\JJ$ on 
$\widehat{\MM}$, a Borel measure $\rho(dm')$ of finite mass on $\widehat{\MM}'$
and a Borel set  
$J(m')$ in $C_r^M$ such that given a continuous function $T$ on 
$\MM_\i$ we have the decomposition
$$
\int T(m)\nu(dm)=\int \rho(dm')\int \vartheta_{m'}(d\xi_M) T(m)
$$
where
$$
\vartheta_{m'}(d\xi_M)=e^{\JJ(m)}1_{J(m')}(\xi_M)d\xi_M
$$
}
\*
\0{\bf Proof:} We show first that the functions $\JJ_N$ converge to a
bounded H\"older continuous function on $\widehat M$.
For this observe that
$$\eqalign{
\l^+(\Psi^N(s,\xi^-,\xi^+))-&\l^+(\Psi^N(s,\xi^-,(0,\xi^\perp)))=\cr
=&\sum_{\bi\in\Omega_N}(\lambda(\tau_\bi\Psi^N( s, \xi^-,
\xi^+))-\l(\tau_\bi\Psi^N( s, \xi^-,
(0,\xi^\perp))))}\Eq(ell)
$$
By the H\"older continuity of $\lambda$  \equ(hold) we have
$$\eqalign{
|\l(\tau_\bi\Psi^N(s,\xi^-,\xi^+))-&\l(\tau_\bi\Psi^N(s,\xi^-,(0,\xi^\perp)))|
\le\cr&
C\e d(\tau_\bi\Psi^N(s,\xi^-,\xi^+),
\tau_\bi\Psi^N(s,\xi^-,(0,\xi^\perp)))^\g
\le\cr&
C\e \left(\sum_\bj 2^{-|\bj|}
|\Psi^N_{\bj-\bi}(s,\xi^-,\xi^+)-\Psi^N_{\bj-\bi}(s,\xi^-,(0,\xi^\perp))|\right)^\g}.
$$
From the regularity property of $\Psi^N$ at fixed $(s,\xi^-)$ we infer
$$
|\Psi^N_\bk(s,\xi^-,\xi^+))-\Psi^N_\bk(s,\xi^-,(0,\xi^\perp))|\le
Ce^{-c dist(\bk,\Omega_M)}\Eq(ell1)
$$
so that
$$
|\l^+(\Psi^N(s,\xi^-,\xi^+))-\l^+(\Psi^N(s,\xi^-,(0,\xi^\perp))|\le
C_M\e
\Eq(esti)
$$
uniformly in $N$. Observe finally that from \equ(esti) we get
$$
|\l^+(\AA^{-i}(\Psi^N(s,\xi^-,\xi^+)))-\l^+
(\AA^{-i}(\Psi^N(s,\xi^-,(0,\xi^\perp)))|\le C_Me^{-ci}\e\Eq(ell3)
$$
because $\Psi^N(s,\xi^-,(0,\xi^\perp))$ and $\Psi^N(s,\xi^-,\xi^+)$
are on the same leave of the unstable foliation.
Convergence follows from convergence of $\lambda$, $\AA$ and $\Psi^N$.
\*
\noindent We will next prove that the masses of the measures
$\rho_N$ are uniformly bounded \ie that $\rho_N(\MM'_N)<C$ with $C$
independent from $N$. 

The set $J_N(s,\xi^-)$ can be
written as

$$
J_N(s,\xi^-)=\bigcap_{\bi\in\Omega_N}
J_{N\bi}(s,\xi^-)
$$
where
$$
J_{N\bi}(s,\xi^-)=\{\xi^+| Y^N_\bi(m)\in U_{s_\bi}\}
$$
where $U_{s}$ is the interval spanning the 
unstable side of the rectangle $Q_s$
of the Markov partition of the linear map $A$. Moreover
$Y^N(m)=X^{-1}(\Psi^N(s,\xi^-,\xi^+))$ is a H\"older continuous function
such that $|\delta_{\xi_M}Y^N_\bi(m)|\le Ce^{-\beta|\bi|}$.

Let us define the functions
$\left(S^\pm(\xi^\perp)\right)_\bi=C^\pm_\bi\xi_\bi^\perp$ where the
$C^\pm_\bi=1\pm Ce^{-\o|\bi|}$ for suitable $C$ and $\o$. We can then
define the two sets

$$K_N^\pm(s,\xi^-)=S^\pm(J_N^\perp(s,\xi^-,0))$$
From the property of the function $Y$ it follows that, for suitable
$C$ and $\o$ we have

$$\eqalign{
J_N^\perp(s,\xi^-,\xi_M)\subset K_N^+(s,\xi^-)\qquad \hbox{for
every}\, \xi_M\in C_r^M\cr
J_N^\perp(s,\xi^-,\xi_M)\supset K_N^-(s,\xi^-)\qquad \hbox{for
every}\, \xi_M\in C_{r/2}^M}
$$

From this follows that
$$
\eqalign{
\int e^{\KK_N(m')} 1_{J^\perp_N(s,\xi^-)}(\xi^\perp)d\xi^\perp&\le \int
e^{\KK_N(m')}1_{K^+_{N}(s,\xi^-)}(\xi^\perp)d\xi^\perp\cr
\int e^{\II_N(m)}1_{J_N(s,\xi^-)}(\xi^+)d\xi^+&\ge e^{-C_\JJ}\int e^{\KK_N(m')}
1_{K^-_{N}(s,\xi^-)}(\xi^-)1_{C^M_{r/2}}(\xi_M)d\xi^\perp d\xi_M}
$$

The following Lemma will allow us to compare the right hand sides of
the two above inequalities.
\*
\0{\bf Lemma\ } {\it For $\xi^\perp\in C_r^{N,M}$ we have
$|\KK_N(s,\xi^-,S^\pm(\xi^\perp))-\KK_N(s,\xi^-,\xi^\perp)|\le C_\KK$.}
\*
\0{\bf Proof}: $\KK_N(s,\xi^-,\xi^\perp)$ is given by eq.\equ(gm). We
can write $$\KK_N(s,\xi^-,\xi^\perp)=\sum_i \KK_{Ni}(s,\xi^-,\xi^\perp).$$
We start bounding the term with $i=0$. 
We have
$$
\eqalign{
|\KK_{N0}(s,\xi^-,S^\pm(\xi^\perp))-&\KK_{N0}(s,\xi^-,\xi^\perp)|\cr&\le
|\l^+(\Psi^N(s,\xi^-,(0,S^\pm(\xi^\perp))))-\l^+
(\Psi^N(s,\xi^-,(0,\xi^\perp)))|\cr&
\le\sum_\bi|\l(\tau_\bi\Psi^N( s, \xi^-,(0,
S^\pm(\xi^\perp))))-\l(\tau_\bi\Psi^N( s, \xi^-,(0, 
\xi^\perp)))|}\Eq(ell2).
$$
We may now proceed as after eq. \equ(ell), replacing eq. \equ(ell1)
by
$$
|\Psi^N_\bk(s,\xi^-,(0,S^\pm(\xi^\perp)))-\Psi^N_\bk(s,\xi^-,(0,\xi^\perp))|\le
Ce^{-c dist(\bk,\Omega_M)}e^{-\o |\bi|}
$$
and bounding \equ(ell2) by $C\e$. As in eq. \equ(ell3)
we the obtain that
$$
|\KK_{Ni}(s,\xi^-,S^\pm(\xi^\perp))-\KK_{Ni}(s,\xi^-,\xi^\perp)|\le
Ce^{-c|i|}\e
$$
which yields the claim.
\*

Using the above Lemma it follows that

$${\int e^{\KK_N(m')} 1_{J^\perp_N(s,\xi^-)}(\xi^\perp)d\xi^\perp\over
\int e^{\II_N(m)}1_{J_N(s,\xi^-)}(\xi^+)d\xi^+}\le C$$

The boundedness of the measure $\rho_N(dm')$ follows from
the above estimate and the fact that $\nu(dm)$ is a probability
measure. 

Let $T(m)$ now be a continuous function on $\hat\MM$. For $m\in\hat\MM$
define $P_N m\in\hat\MM$ to be the point that coincides with $m$ on $\Omega_N$
and is extended periodically outside $\Omega_N$, \ie $P_N m$ is also
in $\hat\MM_N$, see comment before \equ(AN). Set $T_N(m)=T(P_N
m)$. The continuity of $T$ and the weak 
convergence of $\nu_N$ imply
$$
\int T(m)\nu(dm)=\lim_{N\to\infty}\int T_N(m)\nu_N(dm).
$$
Decomposing as in Section 6.2., we get
$$
\int T(m)\nu(dm)=\lim_{N\to\infty}\int b_N(ds)\nu_{Ns}(d\xi^-)\int\nu_{N,s,\xi^-}
(d\xi^+)\widetilde T_N(m)
$$
with $\widetilde  T(m)=e^{\II(m)} T(m)$ and
$\nu_{N,s,\xi^-}=e^{\II_N(m)}1_{J_N(s,\xi^-)}(\xi^+){d\xi^+\over
|J_N(s,\xi^-)|}$.
By the weak convergence of both measures
$$
\int T(m)\nu(dm)=\int b(ds)\nu_{s}(d\xi^-)\lim_{N\to\infty}\int\nu_{N,s,\xi^-}
(d\xi^+)\widetilde T_N(m).
$$
We can rewrite last limit has
$$
\lim_{N\to\infty}\int\nu_{N,s,\xi^-}
(d\xi^+)\widetilde T_N=\lim_{N\to\infty}\int\rho_{N,s,\xi^-}(d\xi^\perp)g_N(m')
$$
where
$$g_N(m')=\int 1_{J_N(m')}(\xi_M)
\widetilde T_N(m) d\xi_M.
$$
Let now
$$
J^K(m')=\bigcap_{N>K} \bigcap_{\bi\in\O_N/\O_K}\bigcap_{\xi_\bi}J_N(\bar m')\Eq(uni)
$$
\ie we take the union over $N\ge K$ and $\bar m'$ such that
$P_K\bar m'= P_K m'$.
Note that $J^K(m')$ depends on $m'$ only through $P_Km$. Set
$$
g^K(m')=\int 1_{J^K(m')}(\xi_M)
\widetilde T_K(m) d\xi_M.
$$
By compactness there is a subsequence of the measures $\rho_{N,s,\xi^-}$
that converges weakly to some $\rho_{s,\xi^-}$. Moreover
$g^K(m')$ are bounded measurable functions
in $C_r^K$, hence they can be approximated by continuous ones on sets
whose complement has arbitrary small Lebesgue measure and thus
arbitrarily small $\rho_{N,s,\xi^-}$ measure, uniformly
in $N$. Hence we get the limit
$$
\lim_{i\to\infty}\int\rho_{N_i,s,\xi^-}(d\xi^+)g^K(m')=
\int\rho_{s,\xi^-}(d\xi^+)g^K(m').
$$
We need to estimate
$$\eqalign{
\int\rho_{N,s,\xi^-}&(d\xi^\perp)((g_N(m')-g^K(m')))
\cr &=\int\rho_{N,s,\xi^-}(d\xi^\perp)\cdot
\int\ll(1_{J_N(m')}(\xi_M)\widetilde T_N(m)-
1_{J^K(m')}(\xi_M)\widetilde T^K(m)\rr)d\xi_M.}
$$
By continuity of $T$, we have that $\|\widetilde T_N-\widetilde T^K\|_\infty\to 0$
as $N,K\to\infty$. Thus it suffices to show
$$
\int\rho_{N,s,\xi^-}(d\xi^\perp)\int\ll(1_{J_N(m')}(\xi_M)-
1_{J^K(m')}(\xi_M)\rr)\widetilde T^K(m)d\xi_M
\Eq(1-1)
$$
tends to zero as $N,K\to\infty$. Since $J_N(m')\subset J^K(m')$
the difference of the characteristic functions is nonzero only if there exists
a $\bj\in\Omega_N$ and $m'$, $\bar m'$, with $P_K\bar m'=P_K m'$, such that
$$
\eqalign{&Y_j(P_N m',\xi_M)\notin U_{s_\bj}
\cr &Y_j(\bar m',\xi_M)\in U_{s_\bj}
}
$$
Moreover, on the support of $\rho_{N,s,\xi^-}$ we have
$$
Y_\bj(P_N m',\tilde\xi_M)\in U_{s_\bj}
$$
for some $\tilde\xi_M$. Recall that $Y_\bj(m)=\xi^+_\bj+\epsilon_\bj(m)$
and for $K\ge |\bj|$ we have
$$
|\epsilon_\bj(P_N m',\xi_M)-\epsilon_\bj(\bar m',\xi_M)|\le
C\e e^{-c(K-|\bj|)}
$$
and for $|\bj|\ge M$ 
$$
|\epsilon_\bj(P_N m',\xi_M)-\epsilon_\bj(P_N m',\tilde\xi_M)|\le
C\e e^{-c(|\bj|-M)|}.
$$
Thus we may conclude that the \equ(1-1) is bounded by
$Ce^{-cK}$ and therefore
$$
\lim_{N\to\infty}\int\rho_{N,s,\xi^-}(d\xi^\perp)g_N(m')
=\lim_{K\to\infty}
\int\rho_{s,\xi^-}(d\xi^+)g^K(m').
$$
Since $P_{K+1}\bar m'= P_{K+1}m'$ implies $P_K\bar m'= P_Km'$
we get $J^{K+1}(m')\subset J^K(m')$. Defining the measurable set
$$
J(m')=\cap_K J^K(m')
$$
we get by
dominated convergence
$$
\lim_{K\to\infty}g^K(m')=\int 1_{J(m')}(\xi_M)
\widetilde T(m) d\xi_M
$$
whereby the proof is completed.

\newsec{The projected SRB measure}

We now turn to the study of the projected SRB measure and to the proof
of Proposition 1 and Theorem 2. We work with general $N\le\infty$
and suppress the $N$-dependence if no confusion can arise.

Recall that $\Psi_0: \MM\to \Tset$ is the the projection to the torus
at the origin of $\Omega$ expressed in the $(s,\xi^-,\xi^+)$ coordinate
representation for $\Psi$.  $\Psi_0$ is continuous on $\MM$ and for
fixed $s,\xi^-$ real analytic in $\xi^+$.
Let $T(\psi)$ be continuous function from $\Tset$ to $\Rset$
and $M<N$ to be fixed later.
By definition of the projection and Proposition 11
$$
\int_{\sTset}T(\psi){\bf P}\mu(d\psi)=\int T(\Psi_0)\mu(d\Psi)=
\int \rho(dm')\int d\xi_M T(\Psi_0(m',\xi_M))
a(m',\xi_M)
\Eq(intg)
$$
where we set
$$
a(m',\xi_M)=e^{\JJ(m)}1_{J(m')}(\xi_M).\Eq(a)
$$
Let $\omega_{m'}(d\psi)$ be the image under $\Psi_0$ of
the measure $l(m',\xi_M)=a(m',\xi_M)d\xi_M$ i.e.
$$
\omega_{m'}(A)=l(\Psi_0(m',\cdot)^{-1}(A)).
$$
Then eq. \equ(intg) may be written as
$$
\int_{\sTset}T(\psi){\bf P}\mu(d\psi)=
\int \rho(dm')\int_{\sTset} \omega_{m'}(d\psi)T(\psi)
\Eq(intg1)
$$
and we need to study next under what conditions the
measure $\omega_{m'}$ is absolutely continuous with
respect to the Lebesgue measure on the torus $\Tset$.

In the $\pm$ coordinates of $\Tset$ we have $\Psi_0=(\psi^+,\psi^-)$
with $\psi^+=\xi_0+{\cal O}(\epsilon)$. 
It will be convenient to change coordinates on $\MM$ by
solving $\xi_0$ in terms of $\psi^+$. Thus
write $\xi_M=(\xi_0,\xi)$
and let $f_{m'\xi}$ be the inverse
of $\xi_0\rightarrow \psi^+(m',\xi_0,\xi)$. Then the map
$\Psi\circ f_{m'\xi}$ provides coordinates $(m',\xi,\psi^+)$
on $\MM$ and in particular
we get for
$\phi=\Psi_0\circ f_{m'\xi}$ 
$$
\phi(m',\xi,\psi^+)=(\psi^+,\psi^-(m',\xi,\psi^+))\Eq(phii)
$$
where $\psi^-$ is continuous in $m'$ and real analytic in
$\xi,\psi^+$. The measure $\omega_{m'}$ is the image of 
$a\circ f_{m'\xi}d\psi^+d\xi$
under the map $\phi$. Our objective is to show that provided
a nondegeneracy condition is satisfied $\omega_{m'}$ is
absolutely continuous with respect to the Lebesgue measure
$d\psi^+d\psi^-$.
Since $a$ is bounded it suffices to show
$\phi(d\psi^+d\xi)$ is absolutely continuous.

Clearly, the absolute continuity fails if the function
$\psi^-$ in eq. \equ(phii) is constant in $\xi$. This
turns out to be both a necessary and a sufficient condition
as we will now set out to prove. 

Let $\xi'=(\xi^\perp,\xi)$ so that $m\in\MM$ is
given by $m=(s,\xi^-,\xi',\psi^+)$. For a multi-index
$\nn=(n_\bi)_{\bi\in\Omega\backslash 0}$ denote
by $|\nn|:= \sum|n_\bi|$ and by ${\rm supp}\; \nn$
the set of $\bi$ s.t. $n_\bi\neq 0$.

\*
\0{\bf Proposition 12\ }
{\it 
Suppose that for some $m\in \MM$, there exists
integer $k\ge 0$ and a multi-index $ \nn\ne 0$ s.t.
$$
\partial_{\psi^+}^k\partial^{ \nn}_{\xi}\phi(m)\neq 0.\Eq(cond0)
$$
Then for every  $m\in \MM$ there exists
$k(m)\ge 0$ and  $ \nn(m)\ne 0$ s.t. \equ(cond0) holds.
Moreover there exists an  integer $M$ such that we may
assume ${\rm supp}\; \nn(m)\subset \Omega_M$ for all $m$.
}

\*
\0{\bf Proof:} Suppose for some $m$ no such $k$ and $\nn$ exist.
By real analyticity of $\phi$ in $\psi^+$ and $\xi$ this means
$\phi(s,\xi^-,\xi,\psi^+)$ is constant in $\xi$ for all $\psi$.
Going back to the coordinates $(s,\xi^-,\xi^+)$ we infer that
the rank of the map $D_{\xi^+}\Psi_0(s,\xi^-,\xi^+)$ is one
for all $\xi^+$ on the domain. 
Since the map $\xi^+\rightarrow\Psi_0(s,\xi^-,\xi^+)$ 
equals the projection $\bf P$ to origin applied to the embedding $S^+_{\Psi'}$
given by Proposition 4, with $\Psi'=\Psi(s,\xi^-,0)$, it follows
that the rank of $D_{\xi^+}{\bf P} S^+_{\Psi'}(\xi^+)$ equals one
for all $\xi^+\in \Rset^{\Omega}$. But the image of $\Rset^{\Omega}$
under  $S^+_{\Psi'}$ is dense in $\TT$ so by continuity the rank
equals one for all $\Psi\in\TT$. This in turn implies that
$\phi(s,\xi^-,\xi,\psi^+)$ is constant in $\xi$ for all $s,\xi^-,\psi$
i.e. the condition \equ(cond0) holds nowhere.
This takes care of the first claim.

The second claim is non-vacuous only for $N=\infty$. Thus
suppose for all $m\in\MM$ $k(m)\ge 0$ and  $ \nn(m)\ne 0$ exist
such that \equ(cond0) holds.
By continuity it holds in a neighborhood of $m$ with the same
$k(m)$ and $\nn(m)$ and thus by compactness of 
$\MM$ we infer  the existence of $M<\i$. 

\*

We continue now the study of the measure $\o_{m'}$ supposing
the condition \equ(cond0) holds. We choose the $M$ in \equ(intg)
as in Proposition 12. Given a point $m=(m',\tilde\xi,\tilde\psi^+)$ 
let us fix $k(m)$ to be the smallest of
the $k$ satisfying \equ(cond0). Then we may write, for $(\xi,
\psi^+)$ in some
neighbourhood $U(m)$ of the origin
$$
\psi^-(m',\tilde\xi+\xi,\tilde\psi^++\psi^+)-\psi^-(m',\tilde\xi,\tilde\psi^+)
=(\psi^+)^{k(m)}f(m',
\xi,\psi^+)
$$
with $f$ real analytic in $(\xi,\psi^+)$ and
$f(m',\xi,0)$  a non constant function.
We need the simple
\*
\0{\bf Lemma 4}: {\it There exist a neighbourhood $V(m)$ of
the origin in $\Rset^{\Omega_M}$
such that the image of the Lebesgue measure
under the map $F:V(m)\rightarrow \Rset^2$ given by 
$(\xi,\psi^+)\rightarrow (\psi^+, (\psi^+)^kf(m',\xi,\psi^+))$
is absolutely continuous with respect to the Lebesgue
measure in $\Rset^2$.} 
\*
{\bf Proof}: see appendix A.
\*

By compactness we may cover $J(m')$ by a finite number of
such neighbourhoods and conclude the absolute continuity of
$\o_{m'}$ for each $m'$:
$$
\o_{m'}(d\psi)=\o_{m'}(\psi)d\psi
$$
with $\o_{m'}(\psi)$ nonnegative and integrable.
Thus \equ(intg1) becomes
$$
\int_{\sTset}T(\psi){\bf P}\mu(d\psi)=
\int \rho(dm')\int_{\sTset} \omega_{m'}(\psi) T(\psi)d\psi.
\Eq(intg2)
$$
Since, by construction, $\int
\o_{m'}(\psi)d\psi\le C$ for all $m'$ 
we can conclude, by the Fubini-Tonelli theorem, that
${\bf P}\mu(d\psi)=\eta(\psi)d\psi$ with
$\eta(\psi)= \int\rho(dm')\omega_{m'}(\psi)$  in $L^1(\Tset)$.
\*
\*
We will now turn to the proof of Proposition 1, i.e.
we will characterize the systems for which the projection is
singular. 
\*\*

\0{\bf Lemma 5:\ } {\it Suppose \equ(cond0) is violated.
Then the unstable manifold  is a product of curves
$$
W^+(\Psi)=\times_{i\in\Omega} \gamma_i(\Psi)
$$
where $\gamma_i(\Psi):\Rset\to\Tset_i$ is an embedding to the 
torus at $i\in\Omega$.}
\*
\0{\bf Proof:} From the proof of Proposition 12 we know that the map $D{\bf P}
S^+_\Psi(\xi)$ has rank
1 for every  $\xi\in\Rset^{\Omega_N}$.
Thus  the vectors $v_i(\Psi, \xi)=\partial_{\xi_i}{\bf P} S^+_\Psi(\xi)
\in \Rset^2$ are parallel. Since $v_0=(0,1)+\OO(\epsilon_0)
\neq 0$ there exist functions  $\lambda_i(\Psi,\xi)$, real analytic in $\xi\in U$,
such that
$$
v_i=\lambda_iv_0.
\Eq(vi)
$$
Let ${\bf P}^+$ be the orthogonal projection in $\Rset^{2\Omega_N}$ to the unstable
space $E^+$ of $\AA_0$ and let 
$$
f_\Psi={\bf P}^+S^+_\Psi.
$$
Since $S^+_\Psi$ is a real analytic embedding in $\Rset^{2\Omega_N}$
and ${\bf P}^+$ is one to one on the image of $S^+_\Psi$ we conclude that
$f_\Psi$ is a real analytic diffeomorphism of $\Rset^{\Omega_N}$. Let
us change the parameterization of $W^+(\Psi)$ using $f_\Psi$, i.e. let
$\tilde S^+_\Psi=S^+_\Psi\circ f_\Psi^{-1}$ and $\tilde
v_i=\partial_{\xi_i}{\bf P}\tilde S^+_\Psi$.
Then 
$$
{\bf P}^+\tilde S^+_\Psi(\xi)=\xi
$$
and hence ${\bf P}^+\tilde v_i=\delta_{i 0}$. On the
other hand by \equ(vi)  
$$
\tilde v_i=
\partial_{\xi_i}{\bf P}\tilde S^+_\Psi=
v_0\circ f_\Psi^{-1}\sum_j\lambda_j\circ f_\Psi^{-1}
\partial_{\xi_i}f_{\Psi j}^{-1} .
$$
Thus combining these identities with ${\bf P}^+v_0\neq 0$ 
we infer that $\sum_j\lambda_j\circ f^{-1}
\partial_{\xi_i}f^{-1}_j=0$. Therefore
$\tilde v_i$ vanishes identically for $i\neq 0$.  
Hence ${\bf P}\tilde S^+_\psi(\xi)$
depends on $\xi$ only through $\xi_0$.

Let $\tau_i$ for $i\in\Omega_N$ be the translation $(\tau_i\Psi)_j=
\Psi_{i+j}$ and on $\xi$  similarly. Then ${\bf P}_i\tilde S^+_{\Psi}(\xi)=
{\bf P}\tilde S^+_{\tau_\bi\Psi}(\tau_\bi\xi)$. Therefore
${\bf P}_i\tilde S^+_{\Psi}(\xi)=\gamma_i(\Psi,\xi_i)$ for a $\gamma_i$
satisfying the claim of the Lemma.

\*\*

Denote by $\OO=X( 0)$ the fixed point of $\AA$. Observe that, due
to the periodic boundary conditions, all components of $\OO$ are equal
to the same value $\psi_\OO$. Thus all the curves $\gamma_i(\OO)$
are identical. 
Since the restriction of $\AA$ to the $\Omega_M$-periodic points of
$\TT$ is $\AA_M$ we may infer that 
$
\times_\Omega W^+_1(\psi_\OO)\subset W^+(\OO)
$. Thus $\gamma_i(\OO)= W^+_1(\psi_\OO)$ and we have obtained
$$
W^+(\OO)=\times_\Omega W^+_1(\psi_\OO)\Eq(prod)
$$
Let $\widetilde\AA=\tilde X^{-1}\AA\widetilde X$ where $\widetilde X=\times_\Omega
X_1$.Then eq. \equ(prod) implies $\widetilde W^+(0)=\times_\bi W^+_A(0)$
where $\widetilde W^+(\Psi)$ and $W^+_A(\psi)$ are the unstable manifolds of the
map $\widetilde\AA$ and of the linear torus map $A$.

Due to the density of $W^+_A(0)$, we get that for any $\Psi\in\TT$
$$
\widetilde W^+(\Psi)=\times_\Omega W^+_A(\Psi_\bi)\Eq(prodbis)
$$
Indeed given $\Psi\in\TT$ we can always find a sequence of points $\Psi_n\in
\widetilde W^+(0)$ such that $\lim_{n\to\i}\Psi_n=\Psi$. Observe that
$\widetilde W^+(\Psi_n)=\times_\Omega W^+_A((\Psi_n)_\bi)$ because
$\widetilde W^+(\Psi_n)=\widetilde W^+(0)$. Let now
$\widetilde W_{r}^+(\Psi_n)$ be the sphere of radius $r$ and center $\Psi_n$ in
$\widetilde W^+(\Psi_n)$. Due to the continuity of the unstable foliation, it
follows that, for every positive $r$, $\widetilde W_{r}^+(\Psi_n)$
converges to $\widetilde W_{r}^+(\Psi)$. This prove \equ(prodbis).

Observe now that, for every $\Psi=(c_\bi e^+_\bi)_{\bi\in\O}\in
\widetilde W^+(0)$ we have that $\widetilde \AA(\Psi)=(c'_\bi
e^+_\bi)_{\bi\in\O}$ where we can write $c'_\bi=\l^+c_\bi+f_\bi(\Psi)$
with $f$ defined and continuous on $\widetilde W^+(0)$. If $\Psi\not
\in \widetilde W^+(0)$ we can again approximate it by a sequence
$\Psi_n$. The continuity of the map $\widetilde \AA$ implies that the
limit of $f(\Psi_n)$ exists and is independent from the chosen
sequence. Finally we obtain

$$
(\widetilde\AA\Psi)_\bi=A\Psi_\bi+f_\bi(\Psi)e_+
$$
which proves our proposition.

\newsec{Perturbative characterization of  singular
couplings}

Proposition 1 gives a geometric characterization of the singular
couplings. This characterization however is not directly testable for
a given interaction $\FF$. We want to discuss here a more practical
although less general way to decide whether a given interaction $\FF$
is singular. For this purpose we will write $\FF=\e\GG$ with
$\GG=O(1)$ and $\e$ small. From proposition 12 and its proof we get immediately that 
\*
\0{\bf Lemma 6} {\it Given $\GG$ if ${\rm rank}(\partial_\e D{\bf P}S^+_\Psi(0)|_{\e=0})
\not \equiv 1$ then there exists $\e_0$, depending
on $\GG$ but not on $N$, such that for all $\e\le\e_0$ the coupled system $\AA_N$,
given by eq.\equ(nu),\equ(trans) with $\FF=\e\GG$, is non degenerate.}
\*
It is rather easy to compute explicitly $\partial_\e D{\bf P}S^+_\Psi(0)|_{\e=0}$ and we get
\*
\0{\bf Lemma 7}: {\it if ${\rm rank}(\partial_\e D{\bf P}S^+_\Psi(0)|_{\e=0})\equiv 1$ then 
$\partial_{e^+_\bi}f^-(\Psi)\equiv 0$}
\*
\0{\bf Proof}: Observe that the first order in $\e$ of the matrix
$\partial_{\xi_\bi}\partial_{\xi_\bj}{\bf P}S^+_\Psi(0)$ 
is the $2\times 2$ matrix obtained by selecting in the $2\O\times \O$
matrix $\chi^+(\Psi)$, see section 4, the rows relative to to the $+$
and $-$ directions of $\Psi_\bo$ and the $\bi$,$ \bj$ columns. 
If ${\rm rank}(\partial_\e D{\bf P}S^+_\Psi(0)|_{\e=0})= 1$ then for every 
$\bi$ and $\bj$ we have ${\rm det}\ll(\partial_\e\partial_{\xi_\bi}
\partial_{\xi_\bj}{\bf P}S^+_\Psi(0)|_{\e=0}\rr)=0$.
By the
choice of the $++$ part of the matrix $\chi^+(\Psi)$, see comment
before \equ(decom), we get that, at first order, 
${\rm det}\partial_{\xi_\bi}\partial_{\xi_\bj}{\bf P}S^+_\Psi(0)=0$ 
for every $f$ unless $\bi=\bo$ or $\bj=\bo$.

Expanding \equ(+++) at first order in $\e$ we get that 
$$
{\rm det}\ll(\partial_\e\partial_{\xi_\bi}\partial_{\xi_\bo}{\bf P}S^+_\Psi(0)|_{\e=0}\rr)=
\ll({\bf T}_1^{-1}\partial_{e^+_\bi} f^-\rr)(\Psi)
$$ 

But  ${\bf T}_1^{-1}$ is a bounded lianear operator, see section 4, so that we must
have $\partial_{e^+_\bi}f^-(\Psi)\equiv 0$ which proves the Lemma.
\*
\*
\0{\bf Acknowledgment:\ } Work supported by NSF Grant DMR-9813268 and
EU grant FMRX-CT98-0175.

\appendices
\newapp{Proof of Lemma 4.}

Suppressing the $m'$ dependence and
denoting $\psi^+$ by $\psi$,
we need to study the image $\eta$ of Lebesgue measure under the map
$$F(\xi,\psi)=(\psi, \psi^kf(\xi,\psi))$$
in some neighbourhood $U$ of the origin of $\Rset^d\times\Rset$.
By assumption, we may write for some $n$ 
$$f(z,0)=\sum_{|\alpha|=n} a_\alpha z^\alpha+\OO(|z|^{n+1})
$$
where not all $a_\alpha$ vanish. Thus 
$
h(z):=\sum_{|\alpha|=n} \tilde a_\alpha z^\alpha
$
is a homogeneous polynomial of degree $n$ that does not vanish
identically and so  there exists $v\in\Rset^d$, $|v|=1$, such that
$h(v)\not=0$. 
Choosing an orthogonal matrix $\OO$ such that $\OO e_d=v$
where $e_d=(0,\ldots,1)$ we see that we may assume without loss
that $a_{(0,\ldots,n)}\not=0$. Writing $z=(u_1,\ldots,u_{d-1},s)$
and defining the function 
$$g(\psi, u, s):= \partial_s f(z,\psi)$$
we may write
$$
g(\psi, u, s)=\sum_r b_r(\psi,u) s^r
$$
so that in a neighbourhood $V$ of the origin of 
$\Rset^{d-1}\times\Rset$
there exists a constant $B$ such that 
$$
|b_r(\psi,u)|\le B^r.
$$
Moreover
$$
\eqalign{b_n(0,0)&=\gamma\not=0\cr
b_r(0,0)&=0\qquad {\rm\ for\ } r<n.\cr}
$$
Chose $\rho>0$ such that 
$$
|b_{n+1}(0,0)s+b_{n+2}(0,0)s^2+\ldots|<\gamma/2
$$
for $|s|\le\rho$. Moreover let $\DD=\{s\in\Cset|\;\; |s|<\rho\}$. Then the
holomorphic function $g(0,0,s)$ has an $n$ fold zero at 0 and no
other zeros in $\DD$. Furthermore
$$
|g(0,0,s)|\ge {|\gamma|\over 2}\rho^n
$$
for $|s|=\rho$. By continuity there is a neighborhood $U$ of zero
in $\Rset^{d-1}\times \Rset$ such that for $(u,\psi)\in U$ 
$$
|g(u,\psi,s)|\ge {|\gamma|\over 4}\rho^n
$$
for $|s|=\rho$. By Rouch\'e's theorem $g(u,\psi,s)$ has exactly $n$
zeros in $\DD$ (counted with multiplicity) when $(u,\psi)\in U$.

Fix $(u,\psi)\in U$ and let $s_1,\ldots,s_m$ be the zeros of
$g(u,\psi,s)$  with multiplicities $n_1,\ldots,n_m$ in $\DD$. 
Then $\prod_i|(s-s_i)^{n_i}|\le
(2\rho)^n$ for $|s|\le\rho$. Therefore 
$$
\phi(s)={g(u,\psi,s)\over \prod_i(s-s_i)^{n_i}}
$$
is analytic in $\DD$, has no zero in $\DD$, and is bounded in
absolute value from below by 
$$
{{|\gamma|\over 4}\rho^n\biggl/ (2\rho)^n}={|\gamma|\over 2^{n+2}}
$$
on $\partial\DD$. By the maximum principle
$$
\phi(s)\ge {|\gamma|\over 2^{n+2}}
$$
for all $s\in\DD$.

Fix now $\psi^+\neq 0$. From the preceeding discussion we infer
that the function $s\rightarrow (\psi^+)^kf((u,s),\psi^+)$
has $m(\psi^+)\le n$ critical points $s_i(\psi^+)$ and
therefore $k\le m(\psi^+)$ critical values $\psi^-_i$.
The function 
$$
\eta_u(\psi^+,\psi^-)=\int ds\delta(\psi^--(\psi^+)^kf((u,s),\psi^+))
$$
is smooth in the complement of these critical values.
Let $\psi^-\in U_i\setminus\psi^-_i$ where $U_i$
is a small enough neighbourhood of $\psi^-_i$. Let
$s_j$ be a critical point giving rise to the critical value $\psi_i^-$.
Integrating over a small neighbourhood $V_j$ of $s_j$ we get
$$
\int_{V_j} ds\delta(\psi^--(\psi^+)^kf((u,s),\psi^+))=
\int_{V_j} ds\delta(\psi^--\psi^-_i-(\psi^+)^k\alpha_j(s)(s-s_j)^{n_j})
$$
where $\alpha_j(s)$ is bounded away from zero in $V_j$. Performing
the integration we obtain
$$
\eta_u(\psi^+,\psi^-)=\sum_j a_j(\psi^-,\psi^+,u)(\psi^--\psi^-_i)^{{1\over
n_j}-1}
$$
where $a_j$ is bounded in $\psi^-\in U_i$ and the sum runs over the
critical points $s_j$ giving rise to the critical value $\psi_i$.
Hence, for each $\psi^+\neq 0$, $\eta_u(\psi^+,\psi^-)$ is integrable
in $\psi^-$ with integral bounded by $1$. Thus, by the Fubini-Tonelli Theorem, it is 
integrable in $(\psi^+,\psi^-)$ and by the same theorem the function
$$
\eta(\psi^+,\psi^-)=\int du\eta_u(\psi^+,\psi^-)
$$
is integrable. It is the density of our measure $\eta$
since the $\eta$ measure of the set $\psi^+=0$ vanishes.
The claim is proved.

\rife{BGM}{BGM}{F. Bonetto, G. Gentile, V. Mastropietro, Electric
field on a surface of constant negative curvature, {\it Ergodic Th. and
Dyn. System}, {\bf 20} (2000) 681-686}

\rife{BK}{BK}{J. Bricmont, A. Kupiainen: High Temperature Expansions and
Dynamical Systems, Commun.Math.Phys. {\bf 178}, 703-732 (1996) 
}

\rife{Bo}{Bo}{R. Bowen: {\it Equilibrium states and the ergodic theory
of Anosov diffeomorphism}, Lecture Notes in Math. {\bf 470}, Springer,
Berlin, (1975)}

\rife{CELS}{CELS}{N. Chernov, G. Eyink, J.E. Lebowitz, Ya.G. Sinai: 
Steady state electric conductivity in the periodic Lorentz gas, 
{\it Commun. in Math.  Phys.} {\bf 154} (1993), 569--601.}

\rife{JP}{JP}{M. Jiang, Y.B. Pesin: Equilibrium measures for coupled map 
lattices: existence uniqueness and finite dimensional approximation
Comm. Math. Phys. {\bf 193}, 675-711 (1998)}

\rife{KYY}{KYY}{P. Frederickson, J. L. Kaplan, E. D. Yorke and J. A.
Yorke, {\it J. Differ.  Equations} {\bf 49} (1983) 183-??.}

\rife{MH}{MH}{B. Moran, W.G. Hoover: Diffusion in the periodic Lorentz
billiard {\it J. Stat. Phys.} {\bf 48}(1987) 709}

\rife{EM}{EM}{D.J. Evans, G.P. Morriss: {\it Statistical Mechanics of Nonequilibrium Liquids} (Accademic, London, 1990)}

\rife{P}{P}{B.L. Holian, W.G. Hoover, H.A. Posh: 
Resolution of Loschmidt's paradox: The origin of 
irreversible behavior in reversible atomistic dynamics, 
Phys. Rev. Lett. {\bf 59}, 10-13 (1987)}

\rife{PS}{PS}{Pesin, Ya.B. Sinai, Ya.G. Pesin: Space-time chaos in the 
system of weakly interacting hyperbolic systems.  J. Geom. Phys.  {\bf 5}  
(1988), 483--492}

\rife{Ru}{Ru}{D. Ruelle: {\it Thermodynamic Formalism},
Addison-Wesley, Reading (Mass.), 1978}

\rife{SRB}{SRB}{G. Gallavotti: Topics in chaotic dynamics, 
Computational physics (Granada, 1994), 271-311, 
Lecture Notes in Phys. {\bf 448}, 
Springer, Berlin, 1995. }

\rife{ST}{ST}{Stochastic modeling}

\rife{Y1}{Y1}{L.-S. Young, {\it Ergod. Th.  and Dynam.  Sys.} {\bf 2}
(1982) 109-??.}

\rife{SS}{SS}{D. Ruelle: Smooth Dynamics and New Theoretical Ideas in 
Nonequilibrium Statistical Mechanics JSP {\bf 95} 393-468 (1999)}

\rife{WW}{WW}{J.L. Lebowitz, G.Eyink: 
Generalized Gaussian Dynamics, Phase-Space Reduction, and
Irreversibility: A Comment. Microscopic Simulations of Complex
Hydrodynamics Phenomena. Proceedings of the NATO Advanced Study Institute,
Alghero, 1991, Plenum, 1992. 
}

\rife{Yar}{Yar}{E. J\"arvenp\"a\"a, M. J\"arvenp\"a\"a}

\rife{Ga}{Ga}{G. Gallavotti, A local fluctuation theorem, 
Phys. A {\bf 263} ,39-50 (1999). }

\rife{BL}{BL}{F. Bonetto, D. Daems, J. L. Lebowitz, and V. Ricci:
Properties of stationary nonequilibrium states in the thermostatted 
periodic Lorentz gas: The multiparticle system
Phys. Rev. E {\bf 65}, 051204 (2002)}

\rife{H}{H}{ W.G. Hoover: {\it Computational statistical mechanics}, Elsevier, 1991.}

\biblio

\bye